\documentclass[aps,
  prd,
  twocolumn,
  showpacs,
  nofootinbib,
  superscriptaddress
]{revtex4-1}

\usepackage{color}
\usepackage{graphicx}
\usepackage{latexsym}
\usepackage{float}
\usepackage{amsmath}
\usepackage{amssymb}
\usepackage{wasysym}
\usepackage{multirow}
\usepackage{epsfig}
\usepackage{hyperref}
\usepackage{subcaption}
\usepackage[export]{adjustbox}
\usepackage[utf8]{inputenc}

\newcommand{\be}{\begin{equation}}
\newcommand{\ee}{\end{equation}}
\newcommand{\ba}{\begin{eqnarray}}
\newcommand{\ea}{\end{eqnarray}}

\begin{document}

\widetext
\leftline{Dated: \today}

\title{Constraints on cosmic strings using data from the first Advanced LIGO observing
run.}

\begin{abstract}
  Cosmic strings are topological defects which can be formed in GUT-scale phase transitions in the early universe. They are also predicted to form in the context of string theory. The main mechanism for a network of Nambu-Goto cosmic strings to lose energy is through the production of loops and the subsequent emission of gravitational waves, thus offering an experimental signature for the existence of cosmic strings. Here we report on the analysis conducted to specifically search for gravitational-wave bursts from cosmic string loops in the data of Advanced LIGO 2015-2016 observing run (O1). No evidence of such signals was found in the data, and as a result we set upper limits on the cosmic string parameters for three recent loop distribution models. In this paper, we initially derive constraints on the string tension $G\mu$ and the intercommutation probability, using not only the burst analysis performed on the O1 data set, but also results from the previously published LIGO stochastic O1 analysis, pulsar timing arrays, cosmic microwave background and Big-Bang nucleosynthesis experiments. We show that these data sets are complementary in that they probe gravitational waves produced by cosmic string loops during very different epochs. Finally, we show that the data sets exclude large parts of the parameter space of the three loop distribution models we consider.
\end{abstract}

\author{%
B.~P.~Abbott,$^{1}$  
R.~Abbott,$^{1}$  
T.~D.~Abbott,$^{2}$  
F.~Acernese,$^{3,4}$ 
K.~Ackley,$^{5}$  
C.~Adams,$^{6}$  
T.~Adams,$^{7}$ 
P.~Addesso,$^{8}$  
R.~X.~Adhikari,$^{1}$  
V.~B.~Adya,$^{9}$
% ...
\\
\medskip
(LIGO Scientific Collaboration and Virgo Collaboration) 
\\
\medskip
{{}$^{*}$Deceased, March 2016. }%
{${}^{\ddag}$Deceased, December 2016. }%
{${}^{\dag}$Deceased, February 2017. }%
{{}$^{\sharp}$Deceased, March 2017. }%
{{}$^{\S}$Deceased, November 2017. }%
}\noaffiliation
\affiliation {LIGO, California Institute of Technology, Pasadena, CA 91125, USA }
\affiliation {Louisiana State University, Baton Rouge, LA 70803, USA }
\affiliation {Universit\`a di Salerno, Fisciano, I-84084 Salerno, Italy }
\affiliation {INFN, Sezione di Napoli, Complesso Universitario di Monte S.Angelo, I-80126 Napoli, Italy }
\affiliation {University of Florida, Gainesville, FL 32611, USA }
\affiliation {LIGO Livingston Observatory, Livingston, LA 70754, USA }
\affiliation {Laboratoire d'Annecy-le-Vieux de Physique des Particules (LAPP), Universit\'e Savoie Mont Blanc, CNRS/IN2P3, F-74941 Annecy, France }
\affiliation {University of Sannio at Benevento, I-82100 Benevento, Italy and INFN, Sezione di Napoli, I-80100 Napoli, Italy }
\affiliation {Albert-Einstein-Institut, Max-Planck-Institut f\"ur Gravi\-ta\-tions\-physik, D-30167 Hannover, Germany }
\affiliation {The University of Mississippi, University, MS 38677, USA }
\affiliation {NCSA, University of Illinois at Urbana-Champaign, Urbana, IL 61801, USA }
\affiliation {University of Cambridge, Cambridge CB2 1TN, United Kingdom }
\affiliation {Nikhef, Science Park, 1098 XG Amsterdam, The Netherlands }
\affiliation {LIGO, Massachusetts Institute of Technology, Cambridge, MA 02139, USA }
\affiliation {Instituto Nacional de Pesquisas Espaciais, 12227-010 S\~{a}o Jos\'{e} dos Campos, S\~{a}o Paulo, Brazil }
\affiliation {Gran Sasso Science Institute (GSSI), I-67100 L'Aquila, Italy }
\affiliation {INFN, Sezione di Roma Tor Vergata, I-00133 Roma, Italy }
\affiliation {Inter-University Centre for Astronomy and Astrophysics, Pune 411007, India }
\affiliation {International Centre for Theoretical Sciences, Tata Institute of Fundamental Research, Bengaluru 560089, India }
\affiliation {University of Wisconsin-Milwaukee, Milwaukee, WI 53201, USA }
\affiliation {Leibniz Universit\"at Hannover, D-30167 Hannover, Germany }
\affiliation {Universit\`a di Pisa, I-56127 Pisa, Italy }
\affiliation {INFN, Sezione di Pisa, I-56127 Pisa, Italy }
\affiliation {OzGrav, Australian National University, Canberra, Australian Capital Territory 0200, Australia }
\affiliation {Laboratoire des Mat\'eriaux Avanc\'es (LMA), CNRS/IN2P3, F-69622 Villeurbanne, France }
\affiliation {LAL, Univ. Paris-Sud, CNRS/IN2P3, Universit\'e Paris-Saclay, F-91898 Orsay, France }
\affiliation {California State University Fullerton, Fullerton, CA 92831, USA }
\affiliation {European Gravitational Observatory (EGO), I-56021 Cascina, Pisa, Italy }
\affiliation {Chennai Mathematical Institute, Chennai 603103, India }
\affiliation {Universit\`a di Roma Tor Vergata, I-00133 Roma, Italy }
\affiliation {Universit\"at Hamburg, D-22761 Hamburg, Germany }
\affiliation {INFN, Sezione di Roma, I-00185 Roma, Italy }
\affiliation {Embry-Riddle Aeronautical University, Prescott, AZ 86301, USA }
\affiliation {Albert-Einstein-Institut, Max-Planck-Institut f\"ur Gravitations\-physik, D-14476 Potsdam-Golm, Germany }
\affiliation {APC, AstroParticule et Cosmologie, Universit\'e Paris Diderot, CNRS/IN2P3, CEA/Irfu, Observatoire de Paris, Sorbonne Paris Cit\'e, F-75205 Paris Cedex 13, France }
\affiliation {Korea Institute of Science and Technology Information, Daejeon 34141, Korea }
\affiliation {West Virginia University, Morgantown, WV 26506, USA }
\affiliation {Center for Gravitational Waves and Cosmology, West Virginia University, Morgantown, WV 26505, USA }
\affiliation {Universit\`a di Perugia, I-06123 Perugia, Italy }
\affiliation {INFN, Sezione di Perugia, I-06123 Perugia, Italy }
\affiliation {Syracuse University, Syracuse, NY 13244, USA }
\affiliation {University of Minnesota, Minneapolis, MN 55455, USA }
\affiliation {SUPA, University of Glasgow, Glasgow G12 8QQ, United Kingdom }
\affiliation {LIGO Hanford Observatory, Richland, WA 99352, USA }
\affiliation {Wigner RCP, RMKI, H-1121 Budapest, Konkoly Thege Mikl\'os \'ut 29-33, Hungary }
\affiliation {Columbia University, New York, NY 10027, USA }
\affiliation {Stanford University, Stanford, CA 94305, USA }
\affiliation {Universit\`a di Camerino, Dipartimento di Fisica, I-62032 Camerino, Italy }
\affiliation {Universit\`a di Padova, Dipartimento di Fisica e Astronomia, I-35131 Padova, Italy }
\affiliation {INFN, Sezione di Padova, I-35131 Padova, Italy }
\affiliation {MTA E\"otv\"os University, ``Lendulet'' Astrophysics Research Group, Budapest 1117, Hungary }
\affiliation {Nicolaus Copernicus Astronomical Center, Polish Academy of Sciences, 00-716, Warsaw, Poland }
\affiliation {University of Birmingham, Birmingham B15 2TT, United Kingdom }
\affiliation {Universit\`a degli Studi di Genova, I-16146 Genova, Italy }
\affiliation {INFN, Sezione di Genova, I-16146 Genova, Italy }
\affiliation {RRCAT, Indore MP 452013, India }
\affiliation {Faculty of Physics, Lomonosov Moscow State University, Moscow 119991, Russia }
\affiliation {SUPA, University of the West of Scotland, Paisley PA1 2BE, United Kingdom }
\affiliation {Caltech CaRT, Pasadena, CA 91125, USA }
\affiliation {OzGrav, University of Western Australia, Crawley, Western Australia 6009, Australia }
\affiliation {Department of Astrophysics/IMAPP, Radboud University Nijmegen, P.O. Box 9010, 6500 GL Nijmegen, The Netherlands }
\affiliation {Artemis, Universit\'e C\^ote d'Azur, Observatoire C\^ote d'Azur, CNRS, CS 34229, F-06304 Nice Cedex 4, France }
\affiliation {Institut de Physique de Rennes, CNRS, Universit\'e de Rennes 1, F-35042 Rennes, France }
\affiliation {Washington State University, Pullman, WA 99164, USA }
\affiliation {Universit\`a degli Studi di Urbino 'Carlo Bo', I-61029 Urbino, Italy }
\affiliation {INFN, Sezione di Firenze, I-50019 Sesto Fiorentino, Firenze, Italy }
\affiliation {University of Oregon, Eugene, OR 97403, USA }
\affiliation {Laboratoire Kastler Brossel, UPMC-Sorbonne Universit\'es, CNRS, ENS-PSL Research University, Coll\`ege de France, F-75005 Paris, France }
\affiliation {Carleton College, Northfield, MN 55057, USA }
\affiliation {Astronomical Observatory Warsaw University, 00-478 Warsaw, Poland }
\affiliation {VU University Amsterdam, 1081 HV Amsterdam, The Netherlands }
\affiliation {University of Maryland, College Park, MD 20742, USA }
\affiliation {Center for Relativistic Astrophysics and School of Physics, Georgia Institute of Technology, Atlanta, GA 30332, USA }
\affiliation {Universit\'e Claude Bernard Lyon 1, F-69622 Villeurbanne, France }
\affiliation {Universit\`a di Napoli 'Federico II', Complesso Universitario di Monte S.Angelo, I-80126 Napoli, Italy }
\affiliation {NASA Goddard Space Flight Center, Greenbelt, MD 20771, USA }
\affiliation {RESCEU, University of Tokyo, Tokyo, 113-0033, Japan. }
\affiliation {OzGrav, University of Adelaide, Adelaide, South Australia 5005, Australia }
\affiliation {Tsinghua University, Beijing 100084, China }
\affiliation {Texas Tech University, Lubbock, TX 79409, USA }
\affiliation {Kenyon College, Gambier, OH 43022, USA }
\affiliation {The Pennsylvania State University, University Park, PA 16802, USA }
\affiliation {National Tsing Hua University, Hsinchu City, 30013 Taiwan, Republic of China }
\affiliation {Charles Sturt University, Wagga Wagga, New South Wales 2678, Australia }
\affiliation {University of Chicago, Chicago, IL 60637, USA }
\affiliation {Pusan National University, Busan 46241, Korea }
\affiliation {The Chinese University of Hong Kong, Shatin, NT, Hong Kong }
\affiliation {INAF, Osservatorio Astronomico di Padova, Vicolo dell'Osservatorio 5, I-35122 Padova, Italy }
\affiliation {INFN, Trento Institute for Fundamental Physics and Applications, I-38123 Povo, Trento, Italy }
\affiliation {Universit\`a di Roma 'La Sapienza', I-00185 Roma, Italy }
\affiliation {Universit\'e Libre de Bruxelles, Brussels 1050, Belgium }
\affiliation {Sonoma State University, Rohnert Park, CA 94928, USA }
\affiliation {Montana State University, Bozeman, MT 59717, USA }
\affiliation {Center for Interdisciplinary Exploration \& Research in Astrophysics (CIERA), Northwestern University, Evanston, IL 60208, USA }
\affiliation {Cardiff University, Cardiff CF24 3AA, United Kingdom }
\affiliation {Universitat de les Illes Balears, IAC3---IEEC, E-07122 Palma de Mallorca, Spain }
\affiliation {The University of Texas Rio Grande Valley, Brownsville, TX 78520, USA }
\affiliation {Bellevue College, Bellevue, WA 98007, USA }
\affiliation {Institute for Plasma Research, Bhat, Gandhinagar 382428, India }
\affiliation {The University of Sheffield, Sheffield S10 2TN, United Kingdom }
\affiliation {California State University, Los Angeles, 5151 State University Dr, Los Angeles, CA 90032, USA }
\affiliation {Universit\`a di Trento, Dipartimento di Fisica, I-38123 Povo, Trento, Italy }
\affiliation {Montclair State University, Montclair, NJ 07043, USA }
\affiliation {National Astronomical Observatory of Japan, 2-21-1 Osawa, Mitaka, Tokyo 181-8588, Japan }
\affiliation {Canadian Institute for Theoretical Astrophysics, University of Toronto, Toronto, Ontario M5S 3H8, Canada }
\affiliation {Whitman College, 345 Boyer Avenue, Walla Walla, WA 99362 USA }
\affiliation {School of Mathematics, University of Edinburgh, Edinburgh EH9 3FD, United Kingdom }
\affiliation {University and Institute of Advanced Research, Gandhinagar Gujarat 382007, India }
\affiliation {IISER-TVM, CET Campus, Trivandrum Kerala 695016, India }
\affiliation {University of Szeged, D\'om t\'er 9, Szeged 6720, Hungary }
\affiliation {Tata Institute of Fundamental Research, Mumbai 400005, India }
\affiliation {INAF, Osservatorio Astronomico di Capodimonte, I-80131, Napoli, Italy }
\affiliation {University of Michigan, Ann Arbor, MI 48109, USA }
\affiliation {American University, Washington, D.C. 20016, USA }
\affiliation {Rochester Institute of Technology, Rochester, NY 14623, USA }
\affiliation {University of Bia{\l }ystok, 15-424 Bia{\l }ystok, Poland }
\affiliation {SUPA, University of Strathclyde, Glasgow G1 1XQ, United Kingdom }
\affiliation {University of Southampton, Southampton SO17 1BJ, United Kingdom }
\affiliation {University of Washington Bothell, 18115 Campus Way NE, Bothell, WA 98011, USA }
\affiliation {Institute of Applied Physics, Nizhny Novgorod, 603950, Russia }
\affiliation {Seoul National University, Seoul 08826, Korea }
\affiliation {Inje University Gimhae, South Gyeongsang 50834, Korea }
\affiliation {National Institute for Mathematical Sciences, Daejeon 34047, Korea }
\affiliation {NCBJ, 05-400 \'Swierk-Otwock, Poland }
\affiliation {Institute of Mathematics, Polish Academy of Sciences, 00656 Warsaw, Poland }
\affiliation {OzGrav, School of Physics \& Astronomy, Monash University, Clayton 3800, Victoria, Australia }
\affiliation {Hanyang University, Seoul 04763, Korea }
\affiliation {NASA Marshall Space Flight Center, Huntsville, AL 35811, USA }
\affiliation {ESPCI, CNRS, F-75005 Paris, France }
\affiliation {Southern University and A\&M College, Baton Rouge, LA 70813, USA }
\affiliation {OzGrav, University of Melbourne, Parkville, Victoria 3010, Australia }
\affiliation {College of William and Mary, Williamsburg, VA 23187, USA }
\affiliation {Indian Institute of Technology Madras, Chennai 600036, India }
\affiliation {IISER-Kolkata, Mohanpur, West Bengal 741252, India }
\affiliation {Scuola Normale Superiore, Piazza dei Cavalieri 7, I-56126 Pisa, Italy }
\affiliation {Universit\'e de Lyon, F-69361 Lyon, France }
\affiliation {Hobart and William Smith Colleges, Geneva, NY 14456, USA }
\affiliation {Janusz Gil Institute of Astronomy, University of Zielona G\'ora, 65-265 Zielona G\'ora, Poland }
\affiliation {University of Washington, Seattle, WA 98195, USA }
\affiliation {King's College London, University of London, London WC2R 2LS, United Kingdom }
\affiliation {Indian Institute of Technology, Gandhinagar Ahmedabad Gujarat 382424, India }
\affiliation {International Institute of Physics, Universidade Federal do Rio Grande do Norte, Natal RN 59078-970, Brazil }
\affiliation {Andrews University, Berrien Springs, MI 49104, USA }
\affiliation {Universit\`a di Siena, I-53100 Siena, Italy }
\affiliation {Trinity University, San Antonio, TX 78212, USA }
\affiliation {Abilene Christian University, Abilene, TX 79699, USA }

\maketitle

\section{Introduction}\label{sec:introduction}
The recent observation of gravitational waves~\cite{PhysRevLett.116.061102} (GWs) has started a new era in astronomy~\cite{Abbott:2017oio,TheLIGOScientific:2017qsa}. In the coming years Advanced LIGO~\cite{0264-9381-32-7-074001} and Advanced Virgo~\cite{0264-9381-32-2-024001} will be targeting a wide variety of GW sources~\cite{Riles20131}. Some of these potential sources could yield new physics and information about the universe at its earliest moments. This would be the case for the observation of GWs from cosmic strings, which are one-dimensional topological defects, formed after a spontaneous symmetry phase transition characterized by a vacuum manifold with non-contractible loops. Cosmic strings were first introduced by Kibble~\cite{Kibble:1976sj}, (for a review see for instance~\cite{Vilenkin:2000jqa,Hindmarsh:1994re,Vachaspati:2015cma}). They can be generically produced in the context of Grand Unified Theories~\cite{Jeannerot:2003qv}. Linear-type topological defects of different forms should leave a variety of observational signatures, opening up a fascinating window to fundamental physics at very high energy scales. In particular, they should lens distant galaxies~\cite{Vilenkin:1984ea,Shlaer:2005ry,Brandenberger:2007ae}, produce high energy cosmic rays~\cite{Vilenkin:1986zz}, lead to anisotropies in the cosmic microwave background~\cite{Kaiser:1984iv,Gott:1984ef}, and produce GWs~\cite{Vachaspati:1984gt,Sakellariadou:1990ne}.

A network of cosmic strings is primarily characterized by the string tension $G\mu$ ($c=1$), where $G$ is Newton's constant and $\mu$ the mass per unit length. The existence of cosmic strings can be tested using the cosmic microwave background (CMB) measurements. Confronting experimental CMB data with numerical simulations of cosmic string networks~\cite{Ade:2013xla,Lizarraga:2016onn,Henrot-Versille:2014jua,Lazanu:2014eya}, the string tension is constrained to be smaller than a few $10^{-7}$.

Cosmic superstrings are coherent macroscopic states of fundamental superstrings (F-strings) and also D-branes extended in one macroscopic direction (D-strings). They are predicted in superstring inspired inflationary models with spacetime-wrapping D-branes~\cite{Sarangi:2002yt,Jones:2003da}. For cosmic superstrings, one must introduce another parameter to account for the fact that they interact probabilistically. In~\cite{Jackson:2004zg}, it is suggested that this intercommutation probability $p$ must take values between $10^{-1}$ and 1 for D-strings and between $10^{-3}$ and $1$ for F-strings. In this paper, we will refer to both topological strings and superstrings as ``strings'', and parameterize them by $p$ and $G\mu$.

Cosmic string parameters can also be accessed through GWs. Indeed, the dynamics of the network is driven by the formation of loops and the emission of GWs. In particular, cusps and kinks propagating on string loops are expected to produce powerful bursts of GWs. The superposition of these bursts gives rise to a stochastic background which can be probed over a large range of frequencies by different observations. Historically, the Big-Bang nucleosynthesis (BBN) data provided the first constraints on cosmic strings~\cite{Cyburt:2004yc}. It was then surpassed by CMB bounds~\cite{Pagano:2015hma} to then be surpassed more recently by pulsar timing bounds~\cite{Lasky:2015lej}. In this paper, we report on the search for GW burst signals produced by cosmic string cusps and kinks using Advanced LIGO data collected between September 12, 2015 06:00 UTC and January 19, 2016 17:00 UTC~\cite{TheLIGOScientific:2016agk}, offering a total of $T_{\textrm{obs}}=4\,163\,421$~s ($\sim48.2$~days) of coincident data between the two LIGO detectors. Moreover, combining the result from the stochastic GW background search previously published in~\cite{TheLIGOScientific:2016dpb}, we test and constrain cosmic string models. While the LIGO O1 burst limit remains weak, the stochastic bound now surpasses the BBN bound for the first time and is competitive with the CMB bound across much of the parameter space.

We will place constraints on the most up-to-date string loop distributions.
In particular, we select three analytic cosmic string models ($M=\{1,2,3\}$)~\cite{Vilenkin:2000jqa, Siemens:2006vk,Blanco-Pillado:2013qja,Lorenz:2010sm,Ringeval:2005kr} for the number density of string loops, developed in part from numerical simulations of Nambu-Goto string networks (zero thickness strings with intercommutation probability equal to unity), in a Friedman-Lema\^{i}tre-Robertson-Walker geometry.
These models are more fully described in Sec.~\ref{sec:models} where their fundamental differences are also discussed.
Sec.~\ref{sec:data} presents an overview of the experimental data sets which are used to constrain the cosmic string parameters. Finally, the resulting limits are discussed in Sec.~\ref{sec:conclusion}.

\section{Cosmic string models}\label{sec:models}

We constrain three different models of cosmic strings indexed by $M$. Common to all these models is the assumption that the width of the strings is negligible compared to the size of the horizon, so that the string dynamics  is given by the Nambu-Goto action.  A further input is the strings intercommutation probability $p$.  
For field theory strings, and in particular $U(1)$ Abelian-Higgs strings in the Bogomol'nyi–Prasad–Sommerfield limit~\cite{Vilenkin:2000jqa}, intercommutation occurs with effectively unit probability \cite{Shellard:1987bv,Laguna:1990it}, $p=1$. That is, when two super-horizon (infinite) strings intersect, they always swap partners; and if a string intersects itself, it therefore chops off a (sub-horizon) loop.  The latter can also result from string-string intersections at two points, leading to the formation of  two new infinite strings and a loop.
  
Cosmic string loops oscillate periodically in time, emitting GWs~\footnote{Super-horizon cosmic strings also emit GWs, due to their small-scale structure~\cite{Sakellariadou:1990ne,Hindmarsh:1990xi,Matsui:2016xnp}.}. A loop of invariant length $\ell$, has period $T=\ell/2$ and corresponding fundamental frequency  $\omega =  4\pi/\ell$. As a result it radiates GWs with frequencies which are multiples of $\omega$, and decays in a lifetime $\tau = \ell/\gamma_{\rm d}$ where~\cite{Casper:1995ub,PhysRevD.50.2496,Vachaspati:1984gt} 
\be
\gamma_{\rm d} \equiv \Gamma G\mu \qquad {\mathrm{with}} \qquad \Gamma \simeq 50\ .
\ee
If a loop contains kinks~\cite{Allen:1991jh,PhysRevD.50.2496,Wachter:2016rwc} (discontinuities on the tangent vector of a string) and cusps (points where the string instantaneously reaches the speed of light), these source bursts of beamed GWs \cite{Damour:2001bk,Damour:2004kw,Siemens:2006yp}. The incoherent superposition of these bursts give rise to a stationary and nearly Gaussian stochastic GW background.
Occasionally, sharp and high-amplitude bursts of GWs stand above this stochastic GW background. 

The three models considered here differ in the loop distribution $n(\ell,t)d\ell$, namely the number density of cosmic string loops of invariant length between $\ell$ and $\ell+d\ell$ at cosmic time $t$. 
To determine the consequences of these differences on their GW signal, 
we work in units of cosmic time $t$ and introduce the dimensionless variables
\be
\gamma\equiv\ell/t \qquad {\mathrm{and}} \qquad {\cal{F}}(\gamma,t) \equiv n(\ell,t) \times t^4.
\label{eq:gammadef}
\ee
We will often refer to $\gamma$ as the relative size of loops and $\cal{F}$ as simply the loop distribution.
All GWs observed today are formed when the string network is in its {\it scaling regime}, namely a self-similar, attractor solution in which all the typical length scales in the problem are proportional to cosmic time~\footnote{Scaling breaks down for a short time in the transition between the radiation and matter eras, and similarly in the transition to dark energy domination.}.

The models considered here were developed (in part) using numerical simulations of Nambu-Goto strings, for which  $p=1$. As mentioned above, cosmic superstrings intercommute with probability $p<1$. The effect of a reduced intercommutation probability on the loop distribution has been studied in~\cite{Avgoustidis:2005nv}. Following this reference we take ${\cal{F}}_{p<1}={\cal{F}}/{p}$~\footnote{In~\cite{Avgoustidis:2005nv} the exponent of the power law behavior was found to be slightly different, namely 0.6. Since our goal here is to highlight the effect of $p<1$, we used a simple dependence of $1/p$ as many others in the litterature have done.}, leading to an increased density of strings~\cite{Sakellariadou:2004wq} and to an enhancement of various observational signatures.

\subsection{Model $M=1$: original large loop distribution}
\label{sec:models:oldLarge}

The first model we consider is the oldest, developed in \cite{Vilenkin:2000jqa,Siemens:2006vk}. It assumes that, in the scaling regime, all loops chopped off the infinite string network are formed with the {\it same} relative size, which we denote by $\alpha$.  At time $t$, the distribution of loops of length $\ell$ to $\ell+d\ell$ contains  loops chopped off the infinite string network at earlier times, and diluted by the expansion of the universe and by the emission of GWs. Assuming that loops do not self-intersect once formed, and taking into account that the length of a loop decays at the rate $d{\ell}/dt=-\gamma_{\rm d}$, the scaling loop distribution (for $\gamma \leq \alpha$) in the radiation era is given by \cite{Vilenkin:2000jqa}
\be
{\cal{F}}_{\rm{rad}}^{(1)}(\gamma) = \frac{C_{\rm rad}}{(\gamma+\gamma_{\rm d})^{5/2}} \Theta(\alpha- \gamma),
\label{dist-rad}
\ee
where $\Theta$ is the Heaviside function, and the superscript $(1)$ stands for model $M=1$.  Some of these loops formed in the radiation era can survive into the matter era, meaning that in the matter era the loop distribution has two components. Those loops surviving from the radiation era have distribution
\ba
{\cal{F}}_{\rm{mat}}^{(1),a} (\gamma,t) &=&   \frac{C_{\rm rad}}{(\gamma+\gamma_{\rm d})^{5/2}} \left(\frac{t_{\rm eq}}{t}\right)^{1/2}\Theta(-\gamma+\beta(t)),
\label{dist-1}
\ea
with $t_{\rm eq}$ the time of the radiation to matter transition, and where the lower bound, $\beta(t)$, is the length in scaling units, of the last loops formed in the radiation era at time $t_{\rm eq}$:
\ba
\beta(t) &=& \alpha \frac{t_{\rm eq}}{t} - \gamma_{\rm d}\left(1-\frac{t_{\rm eq}}{t}\right).
\label{beta}
\ea
The loops formed in the matter era itself have a distribution
\ba
{\cal{F}}_{\rm{mat}}^{(1),b}(\gamma,t) &=& \frac{C_{\rm mat}}{(\gamma+\gamma_{\rm d})^{2}} \Theta(\alpha- \gamma)\Theta(\gamma-\beta(t)).
\label{dist-2}
\ea
The normalisation constants $C_{\rm rad}$ and $C_{\rm mat}$ cannot be determined from analytical arguments, but rather are fixed by matching with numerical simulations of Nambu-Goto strings. Following~\cite{Vilenkin:2000jqa,Siemens:2006vk}: we set them to
\be
C_{\rm rad} \simeq  1.6~~,~~
C_{\rm mat} \simeq  0.48~.
\ee
Furthermore we shall assume that $\alpha \simeq 0.1$. The loop distribution in the matter era is thus given by the sum of distributions in Eqs.~\ref{dist-1} and~\ref{dist-2}.

The loop distribution ${\cal{F}}^{(1)}$ is plotted in Fig.~\ref{fig:loops} for different redshift values and fixing $G\mu$ at $10^{-8}$. A discontinuity, visible for low redshift values, results from the radiation-matter transition which is modeled by Heaviside functions. For $t<t_{eq}$, the loop distribution is entirely determined by Eq.~\ref{dist-rad} and is time independent.


\subsection{Model $M=2$: large loop Nambu-Goto distribution of Blanco-Pillado et al.}

Rather than postulating that all loops are formed with a given size $\alpha t$ at time $t$ as in model 1, the  loop production function can be determined from numerical simulations. This approach was taken in~\cite{Blanco-Pillado:2013qja}, determining the rate of production of loops of size $\ell$ and momentum $\vec{p}$ at time $t$. Armed with this information, $n(\ell,t)$ is determined analytically as in model 1 with the additional assumption that the momentum dependence of the loop production function is weak so that it can be integrated out.  

In the radiation era, the scaling distribution reads
\be
   {\cal{F}}_{\rm{rad}}^{(2)}(\gamma)= \frac{0.18 }{(\gamma+\gamma_{\rm d})^{5/2}}\Theta(0.1 - \gamma),
\label{eq:dist2-1}
\ee
where the superscript $(2)$ stands for model 2. In the matter era, analogously to above, there are two contributions. The loops left over from the radiation era can be deduced from above, whereas loops formed in the matter era have distribution
\begin{equation}
  {\cal{F}}_{\rm{mat}}^{(2),b}(\gamma,t) = \frac{0.27-0.45\gamma^{0.31}}{(\gamma+\gamma_{\rm d})^2} \Theta(0.18 - \gamma)\Theta(\gamma-\beta(t)),
  \label{eq:dist2-2}
\end{equation}
where $\beta(t)$ is given in Eq.~(\ref{beta}) with $\alpha=0.1$.  

The loop distribution of model 2 is plotted in Fig.~\ref{fig:loops}. Notice that in the radiation era, the distributions in models 1 and 2 take the same functional form, though their normalisation differs by a factor of order 10.  In the matter era, the functional form is slightly different and the normalisation is smaller by a factor of order 2.  The authors of~\cite{Blanco-Pillado:2013qja} attribute this reduction in the number of loops to two effects: (i) only about 10\% of the power is radiated into large loops -- indeed, most of it is lost directly into smaller loops which radiate away very quickly; (ii) most of the energy leaving the network goes into loop kinetic energy which is lost to redshifting.

\subsection{Model $M=3$: large loop Nambu-Goto distribution of Ringeval et al.}

This analytical model was presented in~\cite{Lorenz:2010sm}, and is based in part on the numerical simulations of~\cite{Ringeval:2005kr}.  

As opposed to model 2, here the (different) numerical simulation is not used to determine the loop production function at time $t$, but rather the distribution of non-self intersecting loops at time $t$. The analytical modeling also differs from that of model 2 in that an extra ingredient is added:  not only do loops emit GWs --- which decreases their length $\ell$ --- but this GW emission back-reacts on the loops. Back-reaction smooths out the loops on the smallest scales (in particular any kinks), thus hindering the formation of smaller loops~\cite{Wachter:2016hgi,Wachter:2016rwc}. Hence, the distributions of models 2 and 3 differ for the smallest loops.

 Physically, therefore, the model of~\cite{Lorenz:2010sm} contains a further length scale  $\gamma_{\rm c}$, the so-called ``gravitational back-reaction scale'', with 
 $$\gamma_{\rm c} < \gamma_{\rm d},$$
where $\gamma_{\rm d}$ is the gravitational decay scale introduced above. Following the numerical simulation of~\cite{Ringeval:2005kr},
\be
\gamma_{\rm c} = \Upsilon (G\mu)^{1+2\chi}
\ \ \mbox{where}\ \ \Upsilon \sim 10  
\ \ \mbox{and}\ \ \chi = 1-P/2,
\ee
with
\begin{equation}
\label{eq:power-law_p}
\left.
\begin{array}{ccc}
P &= & 1.41\,^{+0.08}_{-0.07} \\
\end{array}
\right|_{\mathrm{mat}},
\quad
\left.
\begin{array}{ccc}
P & = & 1.60\,^{+0.21}_{-0.15} \\
\end{array}
\right|_{\mathrm{rad}}.
\end{equation}

The resulting distribution of loops is given in~\cite{Lorenz:2010sm}. In this paper, we work with the asymptotic expressions given in section 2.4 of ~\cite{Lorenz:2010sm}, valid in the scaling regime ($t\gg t_{\mathrm{ini}}$). Hence the contribution of those loops formed in the radiation era, but which persist into the matter era, are neglected. The loop distribution has three distinct regimes with different power-law behaviours, depending on whether the loops are smaller than $\gamma_c$ ($\gamma \leq \gamma_c$); of intermediate length  ($\gamma_c \leq \gamma \leq \gamma_{\rm d}$); or larger than $\gamma_{\rm d}$ (that is $\gamma_{\rm d} \leq \gamma \leq \gamma_{\rm max}$).  Here $\gamma_{\rm max}=1/(1-\nu)$ is the largest allowed (horizon-sized) loop, in units of cosmic time, where the power-law time evolution of the scale factor of the universe, $a \sim t^\nu$, is
\begin{equation}
\nu=\left. \frac{2}{3} \right|_{\mathrm{mat}}, \qquad \nu=\left. \frac{1}{2} \right|_{\mathrm{rad}}.
\end{equation}
Hence, $\gamma_{\rm max}=2$, or $\gamma_{\rm max}=3$, depending on whether we are in the radiation-dominated or matter-dominated era, respectively.
More explicitly,

\noindent $\bullet$ For loops with length scale large compared to $\gamma_{\rm d}$ :
\begin{equation}
{\cal{F}}^{(3)}(\gamma_{\rm d}\ll \gamma< \gamma_{\rm max})
\simeq \frac{C}{(\gamma+\gamma_{\rm d})^{P+1}}~.
\label{eq:mod3_1}
\end{equation}

\noindent $\bullet$ For loops with length scale 
in the range $\gamma_{\rm c}<\gamma\ll \gamma_{\rm d}$:
\begin{equation}
{\cal{F}}^{(3)}(\gamma_{\rm c}<\gamma\ll \gamma_{\rm d})\simeq 
\frac{C(3\nu-2\chi-1)}{2-2\chi}\frac{1}{\gamma_{\rm d}}\frac{1}{\gamma^{P}}~.
\label{eq:mod3_2}
\end{equation}

\noindent $\bullet$   For loops with length scale smaller than 
$\gamma_{\rm c}$ the distribution is $\gamma$ independent:
\begin{equation}
{\cal{F}}^{(3)}(\gamma\ll \gamma_{\rm c}\ll\gamma_{\rm d})\simeq 
\frac{C(3\nu-2\chi-1)}{2-2\chi}\frac{1}{\gamma_{\rm c}^{P}}\frac{1}{\gamma_{\rm d}}~.
\label{eq:mod3_3}
\end{equation}
Here, $C$ is given by
\ba
C&=&C_0(1-\nu)^{3-P}
\ea
where 
\begin{equation}
\label{eq:power-law_C0}
\left.
\begin{array}{ccc}
C_0 &= & 0.09\,^{-0.03}_{+0.03} \\
\end{array}
\right|_{\mathrm{mat}},
\quad
\left.
\begin{array}{ccc}
C_0 & = & 0.21\,^{-0.12}_{+0.13} \\
\end{array}
\right|_{\mathrm{rad}}.
\end{equation}

In the case of large loops (Eq.~\ref{eq:mod3_1}), $C$ normalizes the distribution. In the radiation era where $\nu=1/2$, $$C\sim 0.08 \qquad \mathrm{(radiation)}$$ (a factor of about 20 smaller than model~1), and in the matter era where $\nu=2/3$, 
$$C\sim 0.016 \qquad \mathrm{(matter)}$$
(a factor of about 30 smaller than model~1).

\begin{figure*}
   \center
   \epsfig{width=5.8cm, file=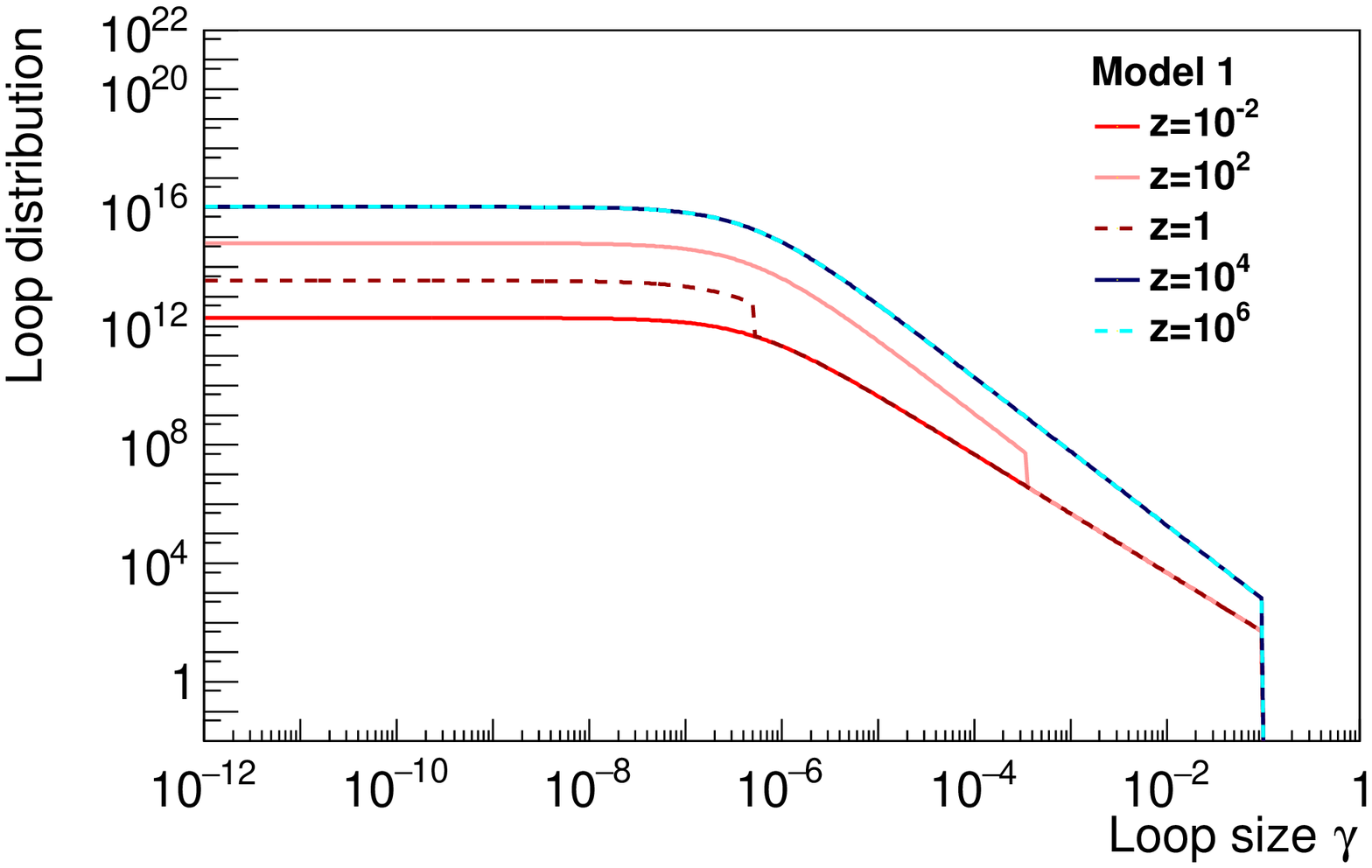}
   \epsfig{width=5.8cm, file=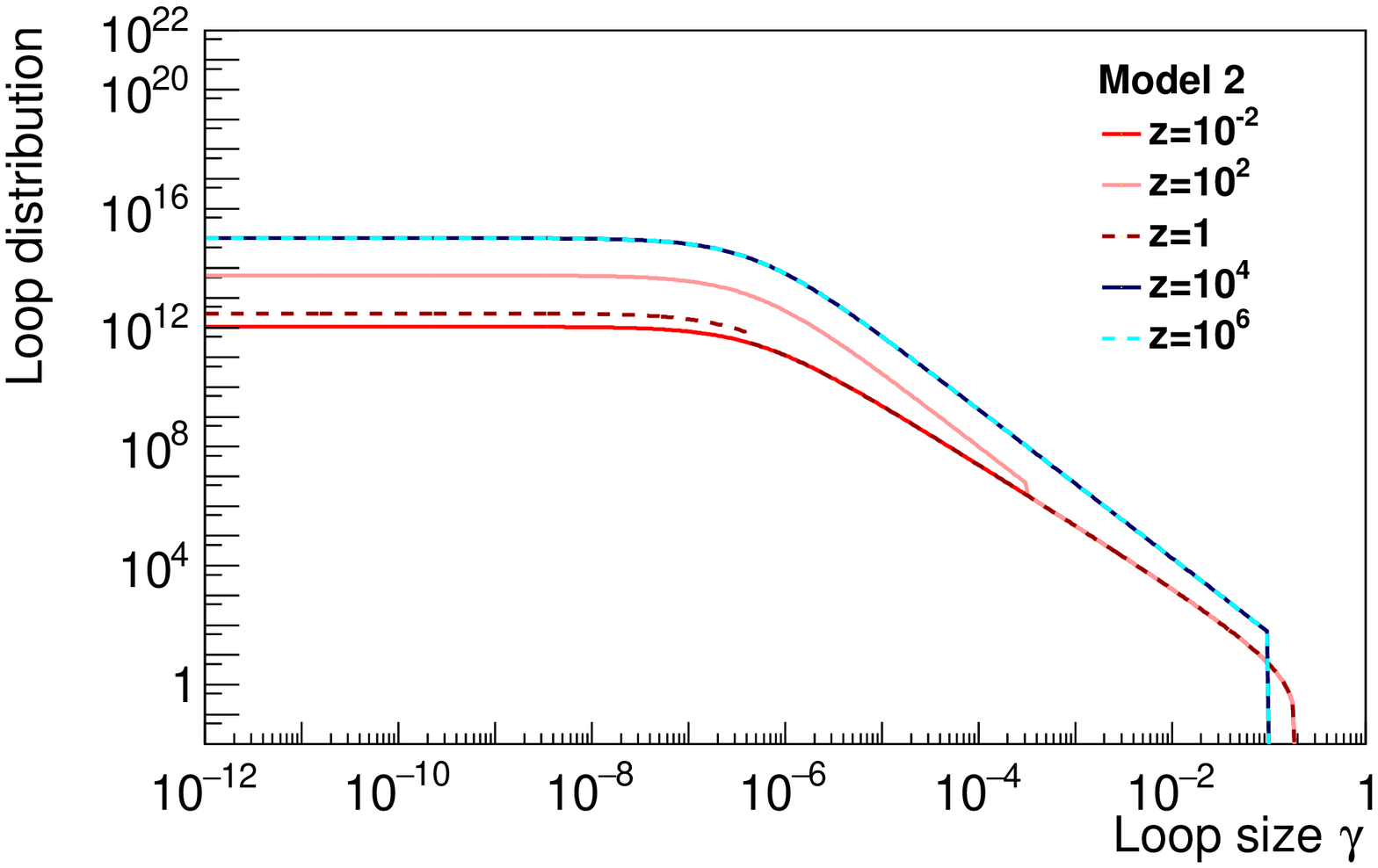}
   \epsfig{width=5.8cm, file=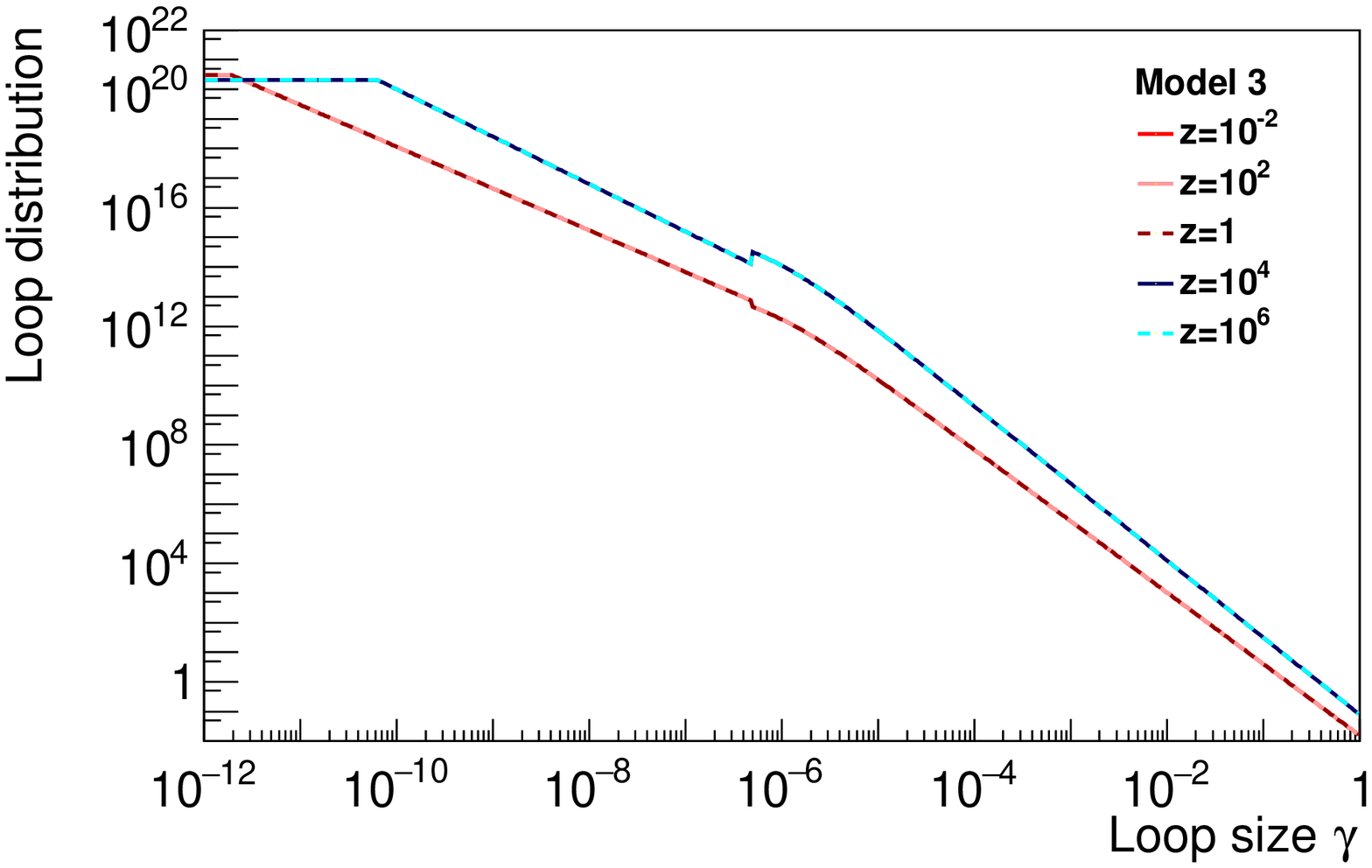}
   \caption{Loop size distributions predicted by three models: $M=1,2,3$. For each model, the loop distribution, ${\cal{F}}(\gamma,t(z))$, is plotted for different redshift values and fixing $G\mu$ at $10^{-8}$.}
   \label{fig:loops}
 \end{figure*}

The three loop regimes are well-visible when plotting the loop distribution: see Fig.~\ref{fig:loops}. Regarding the GW signal, the most significant difference between model 3 and the two previous models is in the very small loop regime ($\gamma \ll \gamma_c$). Comparing Eq.~\ref{eq:mod3_3} with Eq.~\ref{dist-1} and Eq.~\ref{eq:dist2-1}, for models 3, 1 and 2 respectively, in the radiation era, we find
\be
\left. \frac{{\cal{F}}^{(3)}}{{\cal{F}}^{(1,2)}}\right|_{\gamma \ll \gamma_c} \propto (G\mu)^{-0.74}~,\label{eq:small_loops}
\ee
where the proportionality constant is $2.5\times 10^{-2}$ for model 1 and approximately ten times larger for model 2.
 For a typical value of $G\mu = 10^{-8}$, and relative to model $1$, there are $\sim 2\times 10^4$ more very small loops in the radiation era in model 3. As we will see in Sec.~\ref{sec:data}, such a high number of small loops in model 3 will have important consequences in the rate of GW events we can detect and on the amplitude of the stochastic gravitational wave background.

\section{Constraining cosmic strings models with GW data}\label{sec:data}

\subsection{Gravitational waves from cosmic strings}\label{sec:data:gw}

GW bursts are emitted by both cusps and kinks on cosmic string loops, the frequency-domain waveform of which was calculated in~\cite{Damour:2000wa,Damour:2001bk,Damour:2004kw}:
\be
h(\ell,z,f) = A_q(\ell,z) f^{-q} \Theta(f_h-f)\Theta(f-f_\ell),
\label{eq:wf}
\ee
where $q=4/3$ for cusps, $q=5/3$ for kinks, and $A_q(\ell,z)$ is the signal amplitude produced by a cusp/kink propagating on a loop of size $\ell$ at redshift $z$. This waveform is linearily polarized and is only valid if the beaming angle 
\be
\theta_m(\ell,z,f) \equiv (g_2 f (1+z)\ell)^{-1/3} < 1.
\label{thetam}
\ee
Here $g_2$ is an ignorance factor assumed to be 1 in this work (see~\cite{Siemens:2006vk}). In order to detect the GW, the angle subtended by the line of sight and the cusp/kink on a loop of typical invariant length $\ell$ at redshift $z$, must be smaller than $\theta_m$. This condition then determines the high-frequency cutoff $f_h$ in Eq.~\ref{eq:wf}. The low-frequency cutoff $f_\ell$ --- though in principle determined by the kink amplitude, or by the size of the feature that produces the cusp --- is in practice given by the lower end of the GW detector's sensitive band. The amplitude $A_q(\ell,z)$ is given by~\cite{Damour:2001bk}
\be
A_q(\ell,z) = g_1 \frac{G\mu \ell^{2-q}}{(1+z)^{q-1}r(z)}~,
\label{eq:A}
\ee
where the proper distance to the source is given by $r(z) = H_{0}^{-1} \varphi_r(z)$. Here, $H_0$ is the Hubble parameter today and $\varphi_r(z)$ is determined in terms of the cosmological parameters and expressed in Appx.~\ref{appx:cosmo}. Finally $g_1$ gathers together a certain number of uncertainties which enter into the calculation of the cusp and kink waveform (including the amplitude of the cusp/kink, as well as numerical factors of order 1, see~\cite{Damour:2001bk,Siemens:2006vk}). We will set $g_1=1$. In the following, we will use Eq.~\ref{eq:A} to conveniently choose 2 variables out of $\ell$, $z$ and $A_q$. Similarly, we will use Eq.~\ref{eq:wf} to substitute $A_q$ for the strain amplitude $h$.

For a given loop distribution model $M$, in the following we use the GW burst rate derived in~\cite{Siemens:2006vk} and recalled in Appx.~\ref{appx:tech}:
\ba
\frac{d^2R^{(M)}_{q}}{dzdh}(h,z,f) &=& \frac{2N_qH_0^{-3}\varphi_V(z)}{(2-q)(1+z)ht^4(z)} \nonumber \\
&& \times {\cal F}^{(M)}\left(\frac{\ell(hf^q,z)}{t(z)},t(z)\right)  \nonumber \\
&& \times \Delta_{q}(hf^q,z,f).
\label{eq:d2Rdzdh}
\ea
The first two lines on the right-hand side give the number of cusp/kink features per unit space-time volume on loops of size $\ell$, where $N_{q}$ is the number of cusps/kinks per oscillation period $T=\ell/2$ of the loop. In this paper, the number of cusps/kinks per loop oscillation is set to 1 although some models~\cite{Binetruy:2010cc} suggest that this number can be much larger than one. Cosmic time is given by $t(z)=\varphi_t(z)/H_0$ and the proper volume element is $dV(z) = H_0^{-3} \varphi_V(z) dz$ where $\varphi_t(z)$ and $\varphi_V(z)$ are given in Appx.~\ref{appx:cosmo}. Finally $\Delta_q$, which is fully derived in Appx.\ref{appx:tech}, is the fraction of GW events of amplitude $A_q$ that are observable at frequency $f$ and redshift $z$.

\subsection{Gravitational-wave bursts}\label{sec:data:bursts}
We searched the
Advanced LIGO O1 data (2015-2016)~\cite{TheLIGOScientific:2016agk} for individual bursts of GWs from cusps and
kinks. The search for cusp signals was previously conducted using initial LIGO and Virgo data and no signal was found~\cite{Aasi:2013vna}.

For this paper, we use the same analysis pipeline to search for both cusp and kink signals. We perform a Wiener-filter analysis to identify events matching the waveform predicted by the theory~\cite{Damour:2000wa,Damour:2001bk,Damour:2004kw} and given in Eq.~\ref{eq:wf}. GW events are detected by matching the data to a bank of waveforms parameterized by the high-frequency cutoff $f_h$, with $30$~Hz~$<f_h<4096$~Hz. Then resulting events detected at LIGO-Hanford and at LIGO-Livingston are set in time coincidence to reject detector noise artifacts mimicking cosmic string signals. Finally, a multivariate likelihood ratio~\cite{Cannon:2008zz} is computed to rank coincident events and infer probability to be signal or noise. The analysis method is described in~\cite{Aasi:2013vna}. In this paper we only report on the results obtained from the analysis of new O1 LIGO data.

\begin{figure*}
  \center
  \epsfig{width=8cm, file=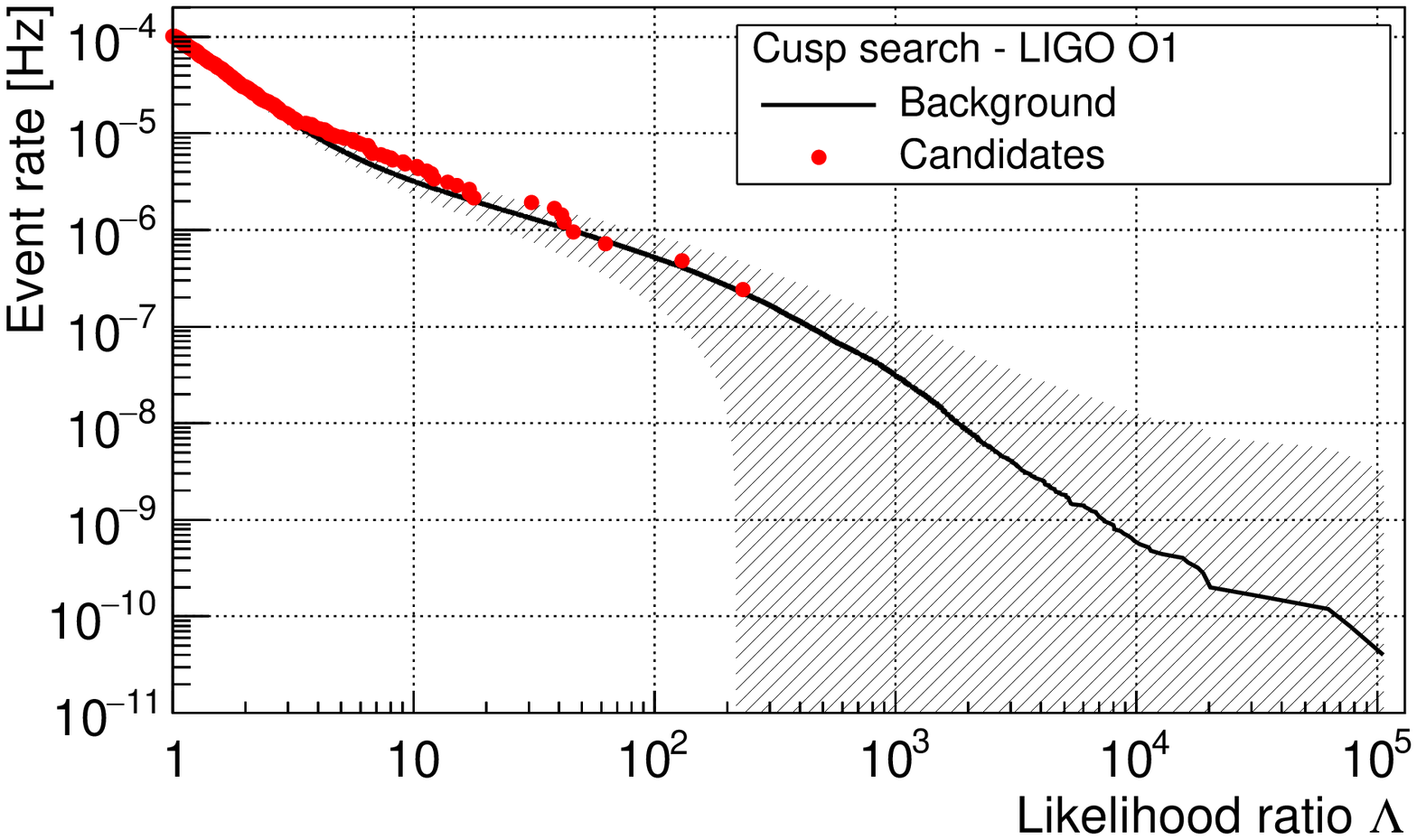} \epsfig{width=8cm, file=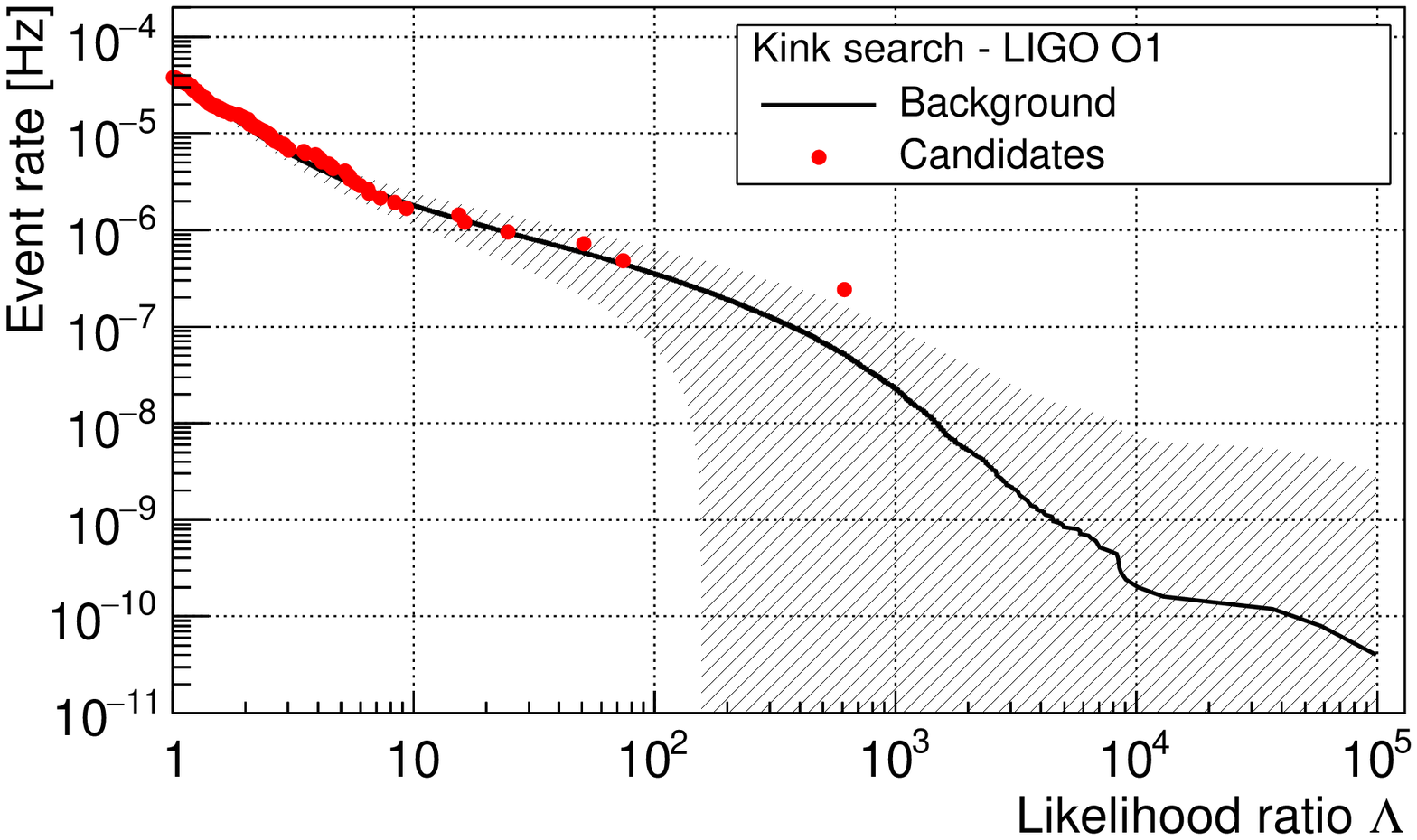}\\
  \epsfig{width=8cm, file=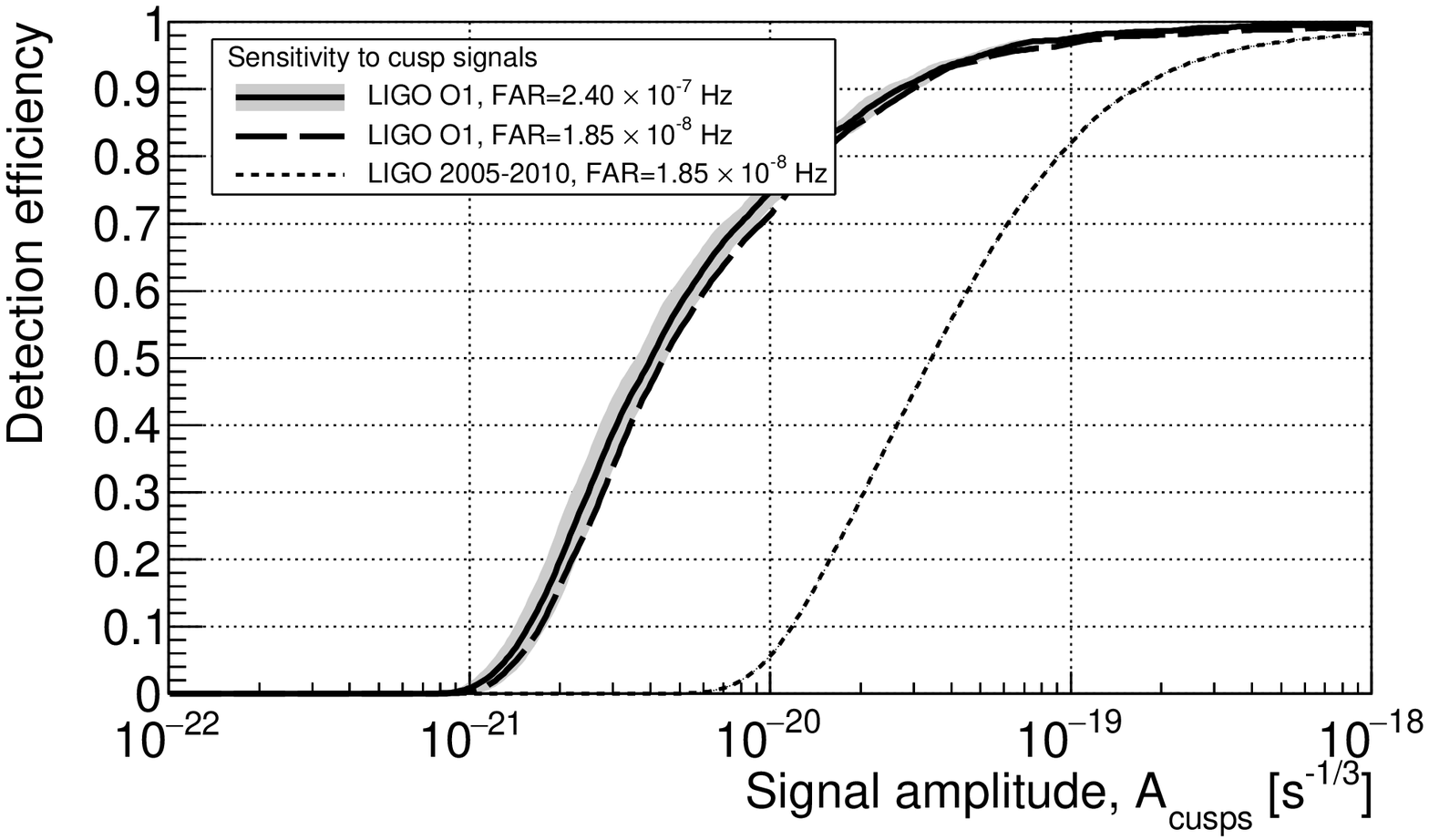} \epsfig{width=8cm, file=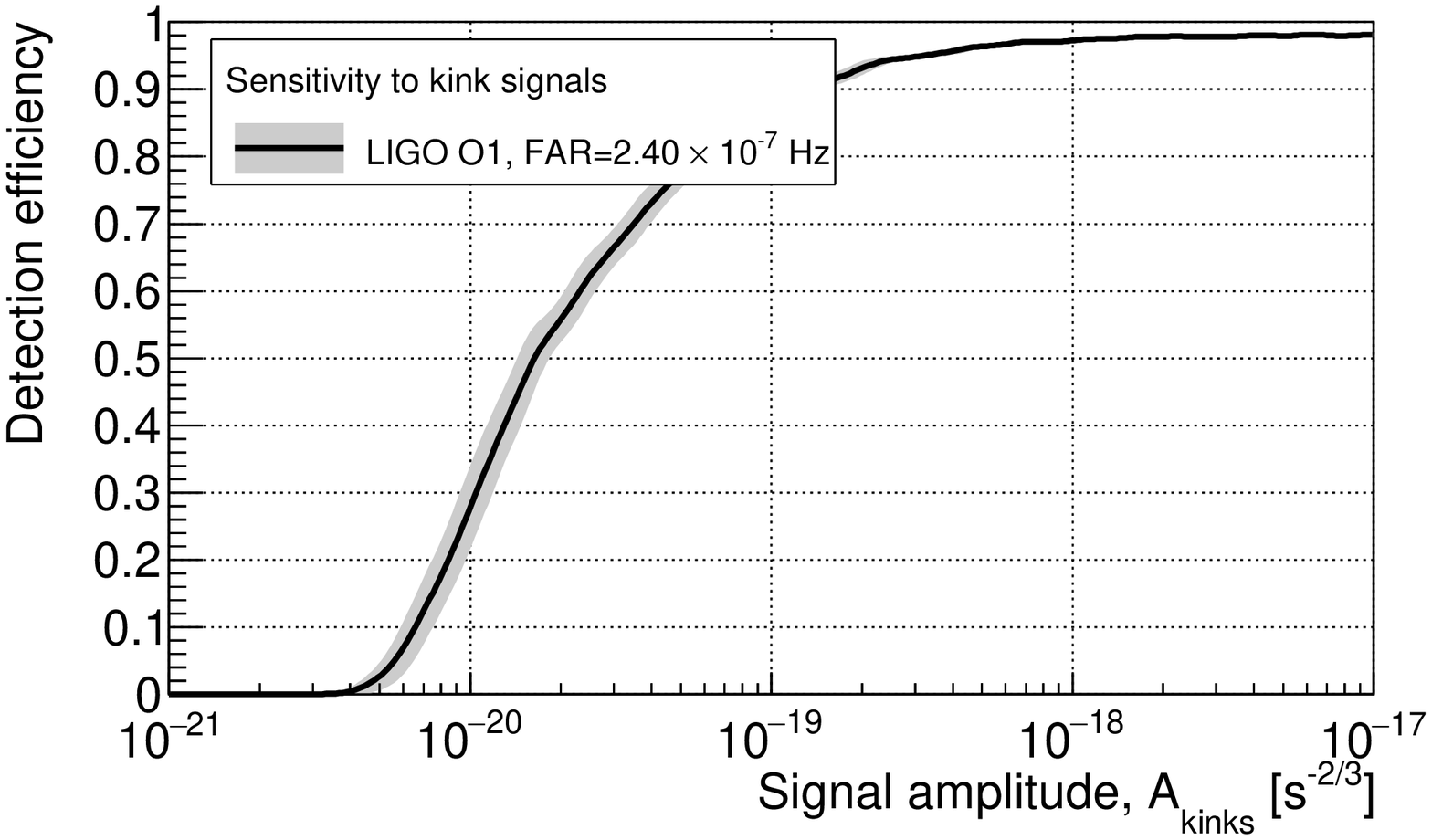}
  \caption{In the upper plots, the red points show the measured cumulative cusp (left-hand plot) and kink (right-hand plot) GW burst rate (using $T_{\rm{obs}}$ as normalization)  as a function of the likelihood ratio $\Lambda$. The black line shows the expected background of the search with the $\pm 1\sigma$ statistical error represented by the hatched area. In both cases, the highest-ranked event ($\Lambda_h \simeq 232$ and $\Lambda_h \simeq 611$) is consistent with the background. The lower plots show the sensitivity of the search as a function of the cusp/kink signal amplitude. This is measured by the fraction of simulated cusp/kink events recovered with $\Lambda > \Lambda_h$. The sensitivity to cusp signals is also measured for a false-alarm rate (FAR) of $1.85 \times 10^{-6}$~Hz to be compared with the sensitivity of the previous LIGO-Virgo burst search~\cite{Aasi:2013vna} (dashed lines).}
  \label{fig:burst}
\end{figure*}

The upper plots in Fig.~\ref{fig:burst} present the final event rate as a function of the likelihood ratio $\Lambda$ for the cusp and kink search. The rate of accidental coincident events between the two detectors (background) is estimated by performing the analysis over 6000 time-shifted LIGO-Livingston data sets. This background data set virtually offers $2.5 \times 10^{10}$~s ($\sim 790.7$~years) of double-coincidence time.  For both cusps and kinks, the candidate ranking values are compatible with the expected background distribution, so no signal was found. The highest-ranked event is measured with $\Lambda_h \simeq 232$ for cusps and $\Lambda_h \simeq 611$ for kinks. These events were scrutinized and were found to belong to a known category of noise transients called ``blips'' described in~\cite{TheLIGOScientific:2016zmo}, matching very well the waveform of cusp and kink signals. 

The sensitivity to cusp and kink GW events is estimated experimentally by injecting simulated signals of known amplitude $A_q$ in the data. We measure the detection efficiency $e_q(A_q)$ as the fraction of simulated signals recovered with $\Lambda > \Lambda_h$, which is associated to a false alarm rate of $1/T_{\rm{obs}}=2.40\times 10^{-7}$~Hz. The detection efficiencies are displayed in the bottom plots in Fig.~\ref{fig:burst}. The sensitivity curve of the 2005-2010 LIGO-Virgo cusp search is also plotted, and should be compared with the O1 LIGO sensitivity measured for an equivalent false-alarm rate of $1.85 \times 10^{-8}$~Hz~\cite{Aasi:2013vna}. The sensitivity to cosmic string signals is improved by a factor 10. This gain is explained by the significant sensitivity improvement at low frequencies of Advanced detectors~\cite{TheLIGOScientific:2016agk}.


Since no signal from cosmic string was found in LIGO O1 data, it is possible to constrain cosmic string parameters using models 1, 2 and 3. To generate statistical statements about our ability to detect true GW signals, we adopt the loudest event statistic~\cite{Brady:2004gt}. We compute an effective detection rate for a given loop distribution model $M$ :
\ba
   {\cal{R}}^{(M)}_q(G\mu,p) &=& \int_0^{+\infty}dA_q\ e_q(A_q) \\
     && \times \int_0^{+\infty}{dz\frac{d^2R^{(M)}_q}{dzdA_q}(A_q,z,f^{*};G\mu,p)},
\label{eq:R}
\ea
where the predicted rate is given by Eq.~\ref{eq:d2Rdzdh} with the change of variables $A_q=hf^{-q}$. The frequency $f^*=30$~Hz is the lowest high-frequency cutoff used in the search template bank as it provides the maximum angle between the line of sight and the cusp/kink on the loop. The parameter space of model $M$, $(G\mu,p)$, is scanned and excluded at a 95\% level when  ${\cal{R}}^{(M)}_q$ exceeds $2.996/T_{\rm{obs}}$ which is the rate expected from a random Poisson process over an observation time $T_{\rm{obs}}$. The resulting constraints are shown in Fig.~\ref{fig:contours} and will be discussed in Sec~\ref{sec:conclusion}.

\subsection{Stochastic gravitational-wave background}\label{sec:data:stochastic}
Cosmic string networks also generate a stochastic background of GWs, which is measured
using the energy density
\be
\Omega_{\rm{GW}}(f) = \frac{f}{\rho_c}\frac{d\rho_{\rm{GW}}}{df},
\ee
where $d\rho_{\rm{GW}}$ is the energy density of GWs in the frequency range $f$ to 
$f + df$ and $\rho_{c}$ is the critical energy density of the Universe. 
Following the method outlined in \cite{Olmez:2010bi}, the GW energy density 
is given by:
\begin{align}
\Omega^{(M)}_{\rm{GW}}(f;G\mu,p) =& \frac{4\pi^2}{3H_0^2}f^3\int_0^{h^{*}}dh\ h^2
\nonumber \\
& \times \int_0^{+\infty}dz \frac{d^2R^{(M)}}{dzdh}(h,z,f;G\mu,p),
\label{eq:spec}
\end{align}
%
%
where the spectrum is computed for a specific 
choice of free parameters $G\mu$ and $p$, and the maximum strain amplitude $h^{*}$ is defined below. 
This equation gives the contribution to the stochastic background from the 
superposition of unresolved signals from cosmic string cusps and kinks, and we shall determine the total GW energy density due to cosmic strings is by summing the two. Note that this calculation {\it underestimates} the stochastic background since it only includes the high-frequency contribution from kinks and cusps. The low-frequency contribution from the smooth part of loops may be important, and has been discussed in \cite{Blanco-Pillado:2017rnf,Blanco-Pillado:2017oxo,Ringeval:2017eww,Binetruy:2012ze}. Neglecting this contribution, conservative constraints will be derived.

To compute the integrals in 
Eq.~\ref{eq:spec} we adopt the numerical method described in 
Appx.~\ref{appx:tech}.
As observed in~\cite{Damour:2000wa}, the 
integration over the strain amplitude is performed up to $h^{*}$ to exclude the individually resolvable 
powerful and rare bursts. The maximum strain amplitude $h^{*}$ determined by solving the equation
\be
\int_{h^{*}}^{+\infty}dh\int_{0}^{+\infty}dz \frac{d^2R^{(M)}}{dzdh}(h,z,f) = f.
\label{eq:hstar}
\ee
This encodes the fact that when the burst rate is larger than $f$, individual bursts are not resolved.

\begin{figure}
  \center
  \epsfig{width=7.7cm, file=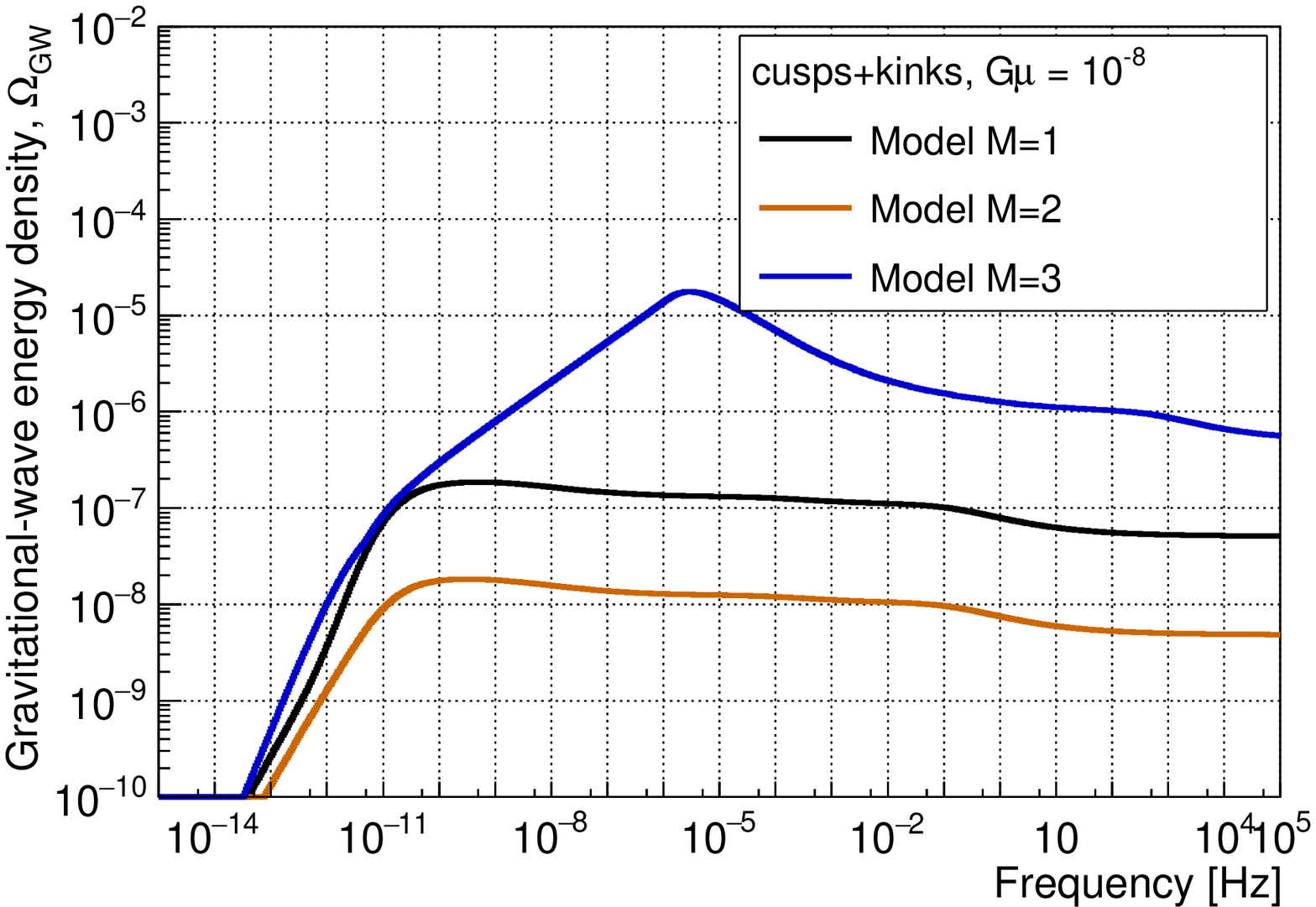}
  \epsfig{width=7.7cm, file=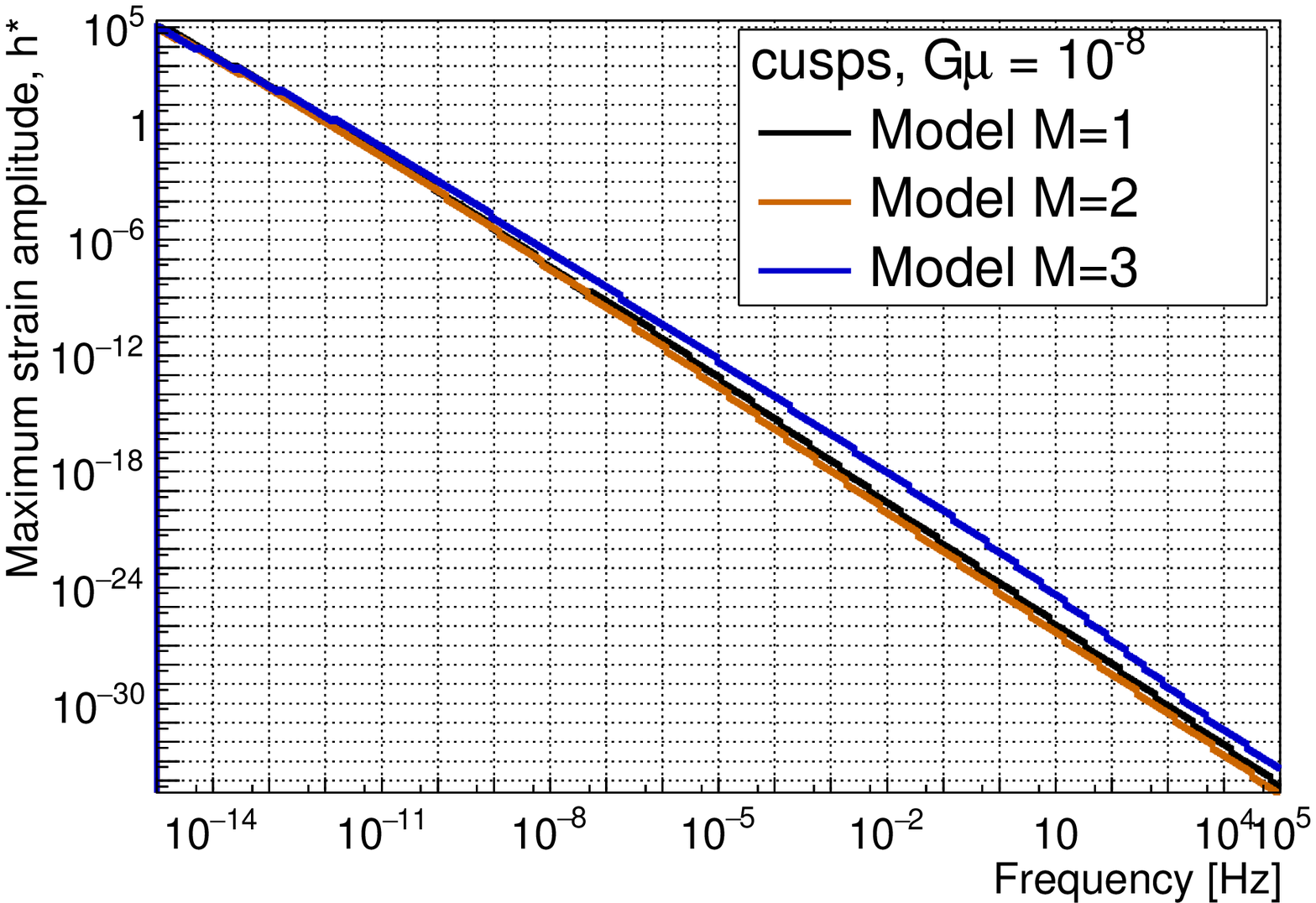}
  \caption{Top: GW energy density, $\Omega^{(M)}_{\rm{GW}}(f)$, from cusps and kinks 
predicted by the three loop distribution models. The string tension $G\mu$ has 
been fixed to $10^{-8}$. Bottom: maximum strain amplitude $h^{*}$ used 
for the integration in Eq.\ref{eq:spec}.}
  \label{fig:spectrum}
\end{figure}

The total energy density in gravitational waves produced by cosmic strings will be composed of overlapping signals ($h < h^{*}$) and non-overlapping signals, namely bursts ($h > h^{*}$). The LIGO-Virgo stochastic search pipeline will detect both types of signals. This has been demonstrated for a stochastic background produced by binary neutron stars, whose signals overlap, and binary black holes, whose signals will arrive in a non-overlapping fashion~\cite{Meacher:2015iua,Abbott:2017xzg}. In this present cosmic string study this effect is negligible: the predicted GW energy density, $\Omega^{(M)}_{\rm{GW}}$,  does not grow significantly (and Fig.~\ref{fig:spectrum}, top, does not change noticeably) when $h^* \rightarrow +\infty$.

\begin{figure}
  \center
  \epsfig{width=7.7cm, file=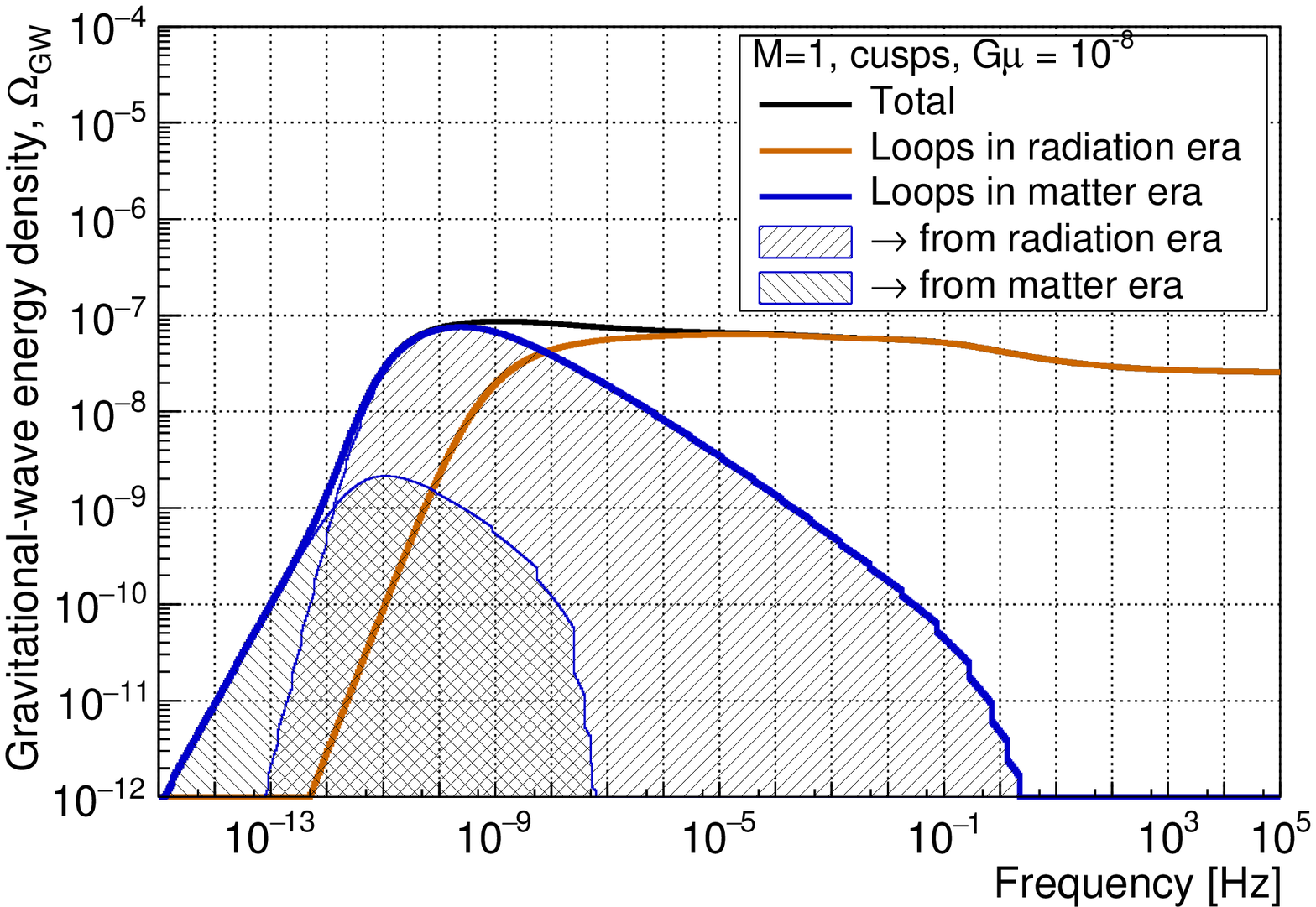}
  \epsfig{width=7.7cm, file=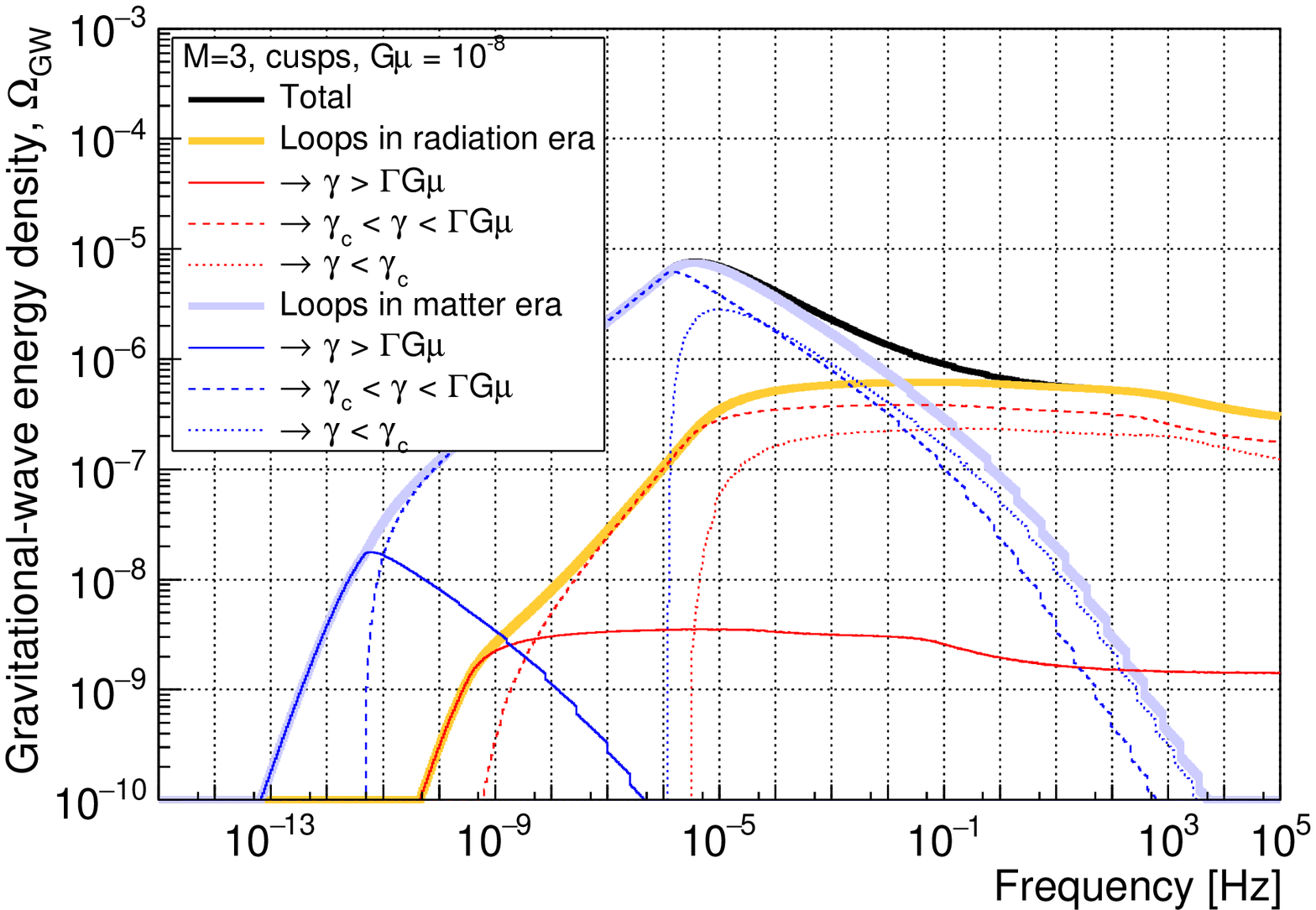}
  \caption{Top: GW energy density, $\Omega^{(M)}_{\rm{GW}}(f)$, from cusps for model 1. We have separated the contributions 
from loops in the radiation ($z>3366$) and matter ($z<3366$) eras. 
Additionally, for loops in the matter era, we have separated the effect 
of loops produced in the matter era from the ones produced in the 
radiation era (Eq.~\ref{dist-rad}, Eq.~\ref{dist-1} and Eq.~\ref{dist-2}). 
Bottom: GW energy density, $\Omega^{(M)}_{\rm{GW}}(f)$, from cusps for model 3. The effect of the three loop 
size regimes is shown (Eq.~\ref{eq:mod3_1}, Eq.~\ref{eq:mod3_2} and 
Eq.~\ref{eq:mod3_3}) for the matter and radiation eras.}
  \label{fig:spectrum_03}
\end{figure}

Fig.~\ref{fig:spectrum} (top) shows the spectra for the three models under 
consideration,
adding both the cusp and the kink contributions and assuming $G\mu=10^{-8}$. 
Model 2 spectrum is about 10 times weaker than the spectrum of model 1 
over most of the frequency range. As shown in Fig.~\ref{fig:spectrum_03} (top),
the spectra are dominated by the contribution of loops in the radiation era 
over most of the frequency range,
including the frequencies accessible to LIGO and Virgo detectors (10-1000 Hz).
The difference in normalizations
of the loop distributions in the radiation era in the two models, 
discussed in Sec.~\ref{sec:models}, is therefore
the cause for the difference in spectral amplitudes. Note also that at
low frequencies ($\sim 10^{-9}$ Hz), at which pulsar timing observations 
are made, the matter era loops contribute the most.

Fig.~\ref{fig:spectrum} (top) also shows that the spectrum for model 3 has a significantly higher amplitude than those of models 1 and 2. Fig.~\ref{fig:spectrum_03} shows that this spectrum is dominated by the contribution of small loops which, as discussed in Sec.~\ref{sec:models}, are much more numerous in model 3.

Fig.~\ref{fig:spectrum} (bottom) shows the maximum value for the 
strain amplitude to consider in the integration, $h^*$ as a 
function of the frequency. At LIGO-Virgo frequencies (10-1000 Hz) the 
spectrum originates from GWs with strain amplitudes below $\sim 10^{-28}$.


The energy density spectra predicted by the models can be compared with several
observational results. First, searches for the stochastic GW 
background using LIGO and Virgo detectors have been performed, using
the initial generation detectors (science run S6, 2009-2010)~\cite{Aasi:2014zwg} and the first 
observation run (O1, 2015-2016) of the advanced
detectors~\cite{TheLIGOScientific:2016dpb}. Both searches reported frequency-dependent
upper limits on the energy density in GWs. To translate these
upper limits into constraints on cosmic string parameters, we define the 
following likelihood function:
\begin{equation}
\ln L(G\mu,p)\propto \sum_i \frac{-\left( Y(f_i) - \Omega^{(M)}_{\rm{GW}}(f_i; G\mu,p)
\right)^2}{\sigma^2(f_i)} ,                                                
\label{stochlik}
\end{equation}                                                                       %
where $Y(f_i)$ and $\sigma(f_i)$ are the measurement and the associated 
uncertainty of the GW energy density in the frequency bin $f_i$, 
and $\Omega_{\rm{GW}}^{(M)}(f_i;G\mu,p)$ is the energy density computed by a cosmic 
string model at the same frequency bin $f_i$ and for some set of model 
parameters $G\mu$ and $p$. We evaluate the likelihood function across 
the parameter space $(G\mu,p)$ and compute the 95\% confidence contours for 
the initial LIGO-Virgo (S6, $41.5<f<169$~Hz)~\cite{Aasi:2014zwg} and for the most 
recent Advanced LIGO (O1, $20<f<86$~Hz)~\cite{TheLIGOScientific:2016dpb} stochastic 
background measurements (assuming Bayesian formalism and 
flat priors in the log parameter space). 
Since a stochastic background of GWs has not been detected yet, these 
contours define the excluded regions of the parameter space. We also compute 
the projected design sensitivity for the Advanced LIGO and Advanced Virgo 
detectors, using Eq.~\ref{stochlik} with $Y(f_i) = 0$ and with the 
projected $\sigma(f_i)$ for the detector network \cite{TheLIGOScientific:2016wyq}.

Another limit can be computed based on the Pulsar Timing Array (PTA) 
measurements of the pulse arrival times of millisecond pulsars
~\cite{Lasky:2015lej}. This measurement produces a limit on the energy
density at nanohertz frequencies --- specifically, at 95\% confidence
$\Omega_{\rm{GW}}^{\rm{PTA}}(f=2.8\times 10^{-9}\mathrm{~Hz})<2.3\times 10^{-10}$. We directly
compare the spectra predicted by our models (at $2.8\times 10^{-9}\mathrm{~Hz}$)
to this constraint.

Finally, indirect limits on the total (integrated over frequency) energy 
density in GWs can be placed based on the Big-Bang Nucleosynthesis
(BBN) and Cosmic Microwave Background (CMB) observations. The BBN model and 
observations of the abundances of the lightest nuclei can be used to
constrain the effective number of relativistic degrees of freedom at the time
of the BBN, $N_{\mathrm{eff}}$. Under the assumption that only photons and 
standard light neutrinos contribute to the radiation energy density, 
$N_{\mathrm{eff}}$ is equal to the effective number of neutrinos, corrected 
for the residual heating of the neutrino fluid due to electron-positron 
annihilation: $N_{\mathrm{eff}} \simeq 3.046$~\cite{Mangano:2005cc}. 
Any deviation from this value can be attributed to extra relativistic radiation,
including potentially GWs due to cosmic string kinks and cusps
generated prior to BBN. We therefore use the 95\% confidence upper limit
$N_{\mathrm{eff}} - 3.046 < 1.4$, obtained by comparing the BBN model and the 
abundances of deuterium and $^4{\rm He}$ ~\cite{Cyburt:2004yc}, which translates into
the following limit on the total energy density in GWs:
\begin{equation}
\Omega^{\rm{BBN}}_{\rm{GW}} (G\mu,p)  =  \int_{10^{-10}{\rm ~Hz}}^{10^{10} {\rm ~Hz}}df
\Omega_{\rm{GW}}^{(M)}(f;G\mu,p) < 1.75 \times 10^{-5},
\label{om_bbn}
\end{equation}
where the lower bound on the integrated frequency region is determined by 
the size of the horizon at the time of BBN~\cite{Binetruy:2012ze}.
In this calculation we only consider kinks and cusps generated before 
BBN, which implies limiting the redshift integral in Eq.~\ref{eq:spec} to
$z>5.5\times 10^{9}$.

Similarly, presence of GWs at the time of photon decoupling
could alter the observed CMB and Baryon Acoustic Oscillation spectra. We apply
a similar procedure as in the BBN case, integrating over redshifts before
the photon decoupling ($z>1089$) and over all frequencies above $10^{-15}$ Hz
(horizon size at the time of decoupling) to compute
the total energy density of GWs at the time of decoupling.
We then compare this quantity to the posterior 
distribution obtained in~\cite{Pagano:2015hma} to compute the 95\%
confidence contours:
\begin{equation}
\Omega^{\rm{CMB}}_{\rm{GW}} (G\mu,p)  =  \int_{10^{-15}{\rm ~Hz}}^{10^{10} {\rm ~Hz}}df
\Omega_{\rm{GW}}^{(M)}(f;G\mu,p) < 3.7 \times 10^{-6},
\label{om_cmb}
\end{equation}
%

\begin{figure}
  \center
  \epsfig{width=8cm, file=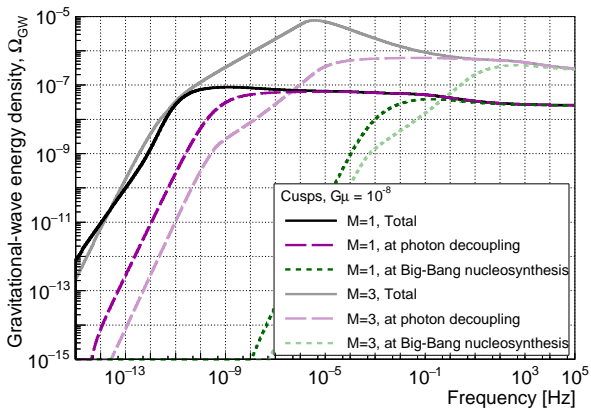}
  \caption{GW energy density, $\Omega^{(M)}_{\rm{GW}}(f)$, from cusps for models 1 and 3. The spectra have 
been computed at the time of photon decoupling ($z_\mathrm{CMB}=1100$) and at 
the time of nucleosynthesis ($z_\mathrm{BBN}=5.5 \times 10^9$).}
  \label{fig:spectrum_ccncmb}
\end{figure}

For reference, Fig.~\ref{fig:spectrum_ccncmb} shows the energy density spectra
for models 1 and 3 using $G\mu=10^{-8}$. As expected, 
the contribution from the matter era loops is suppressed at the time of the BBN
or of photon decoupling, resulting in the suppression of the spectra at low 
frequencies. To have negligible systematic errors associated to the numerical integration, we compute Eq.~\ref{om_bbn} and Eq.~\ref{om_cmb} using 200 and 250 logarithmically-spaced frequency bins respectively.

\begin{figure*}[!tp]
  \centering
  \begin{subfigure}[t]{0.48\textwidth}
    \includegraphics[width=\textwidth]{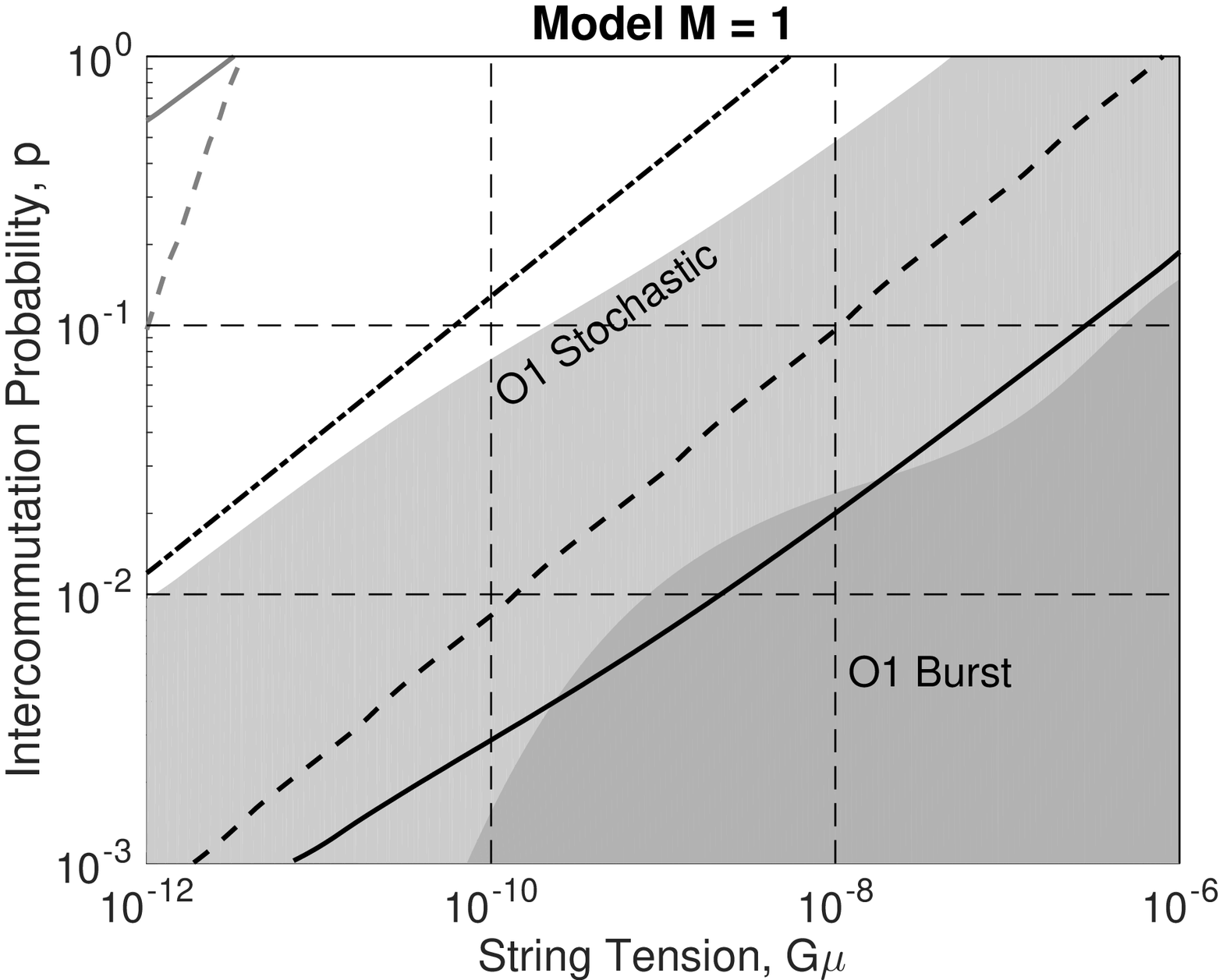}
  \end{subfigure}
  \begin{subfigure}[t]{0.48\textwidth}
    \includegraphics[width=\textwidth]{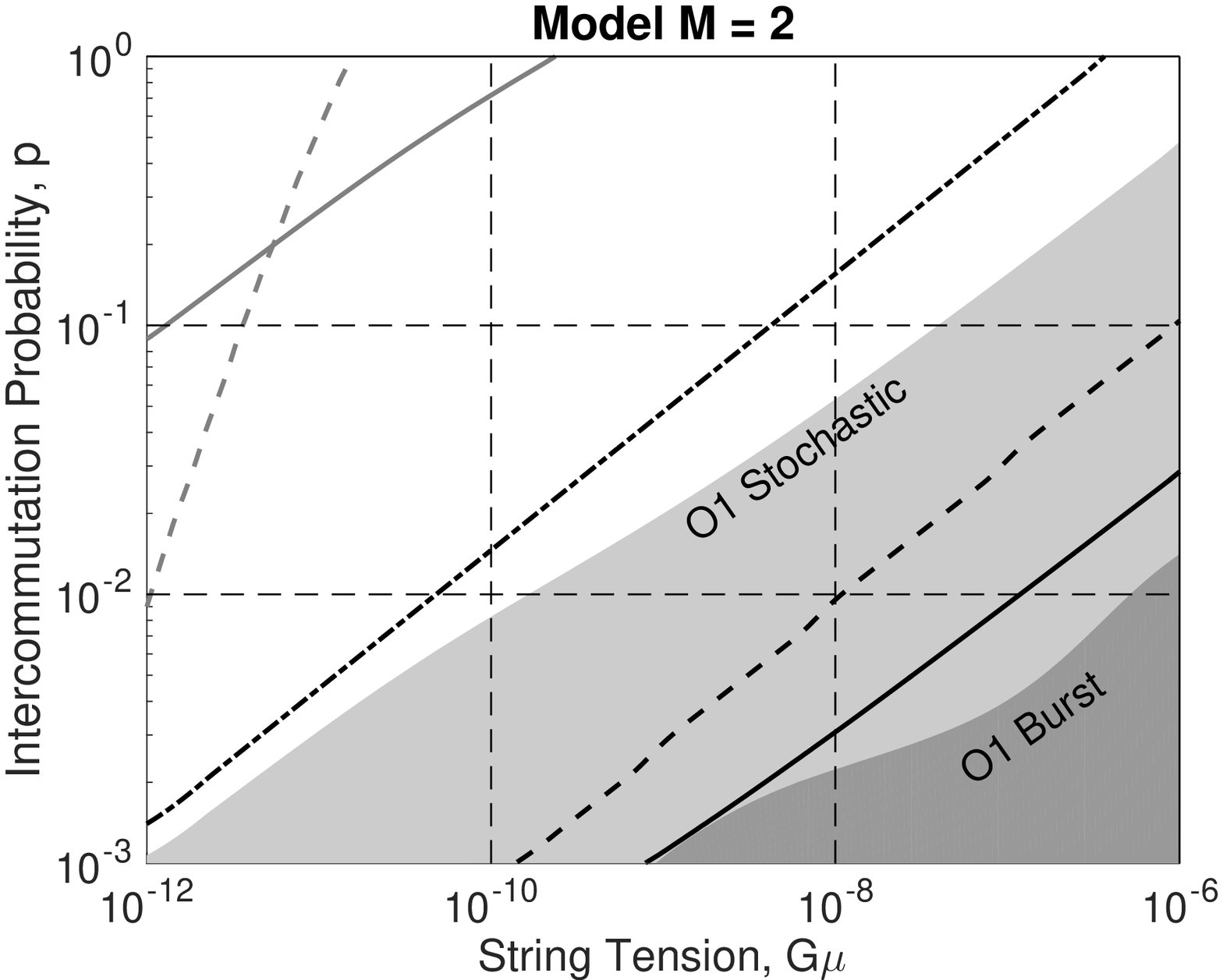}
  \end{subfigure}
  \begin{subfigure}[t]{0.48\textwidth}
    \includegraphics[width=\textwidth,valign=c]{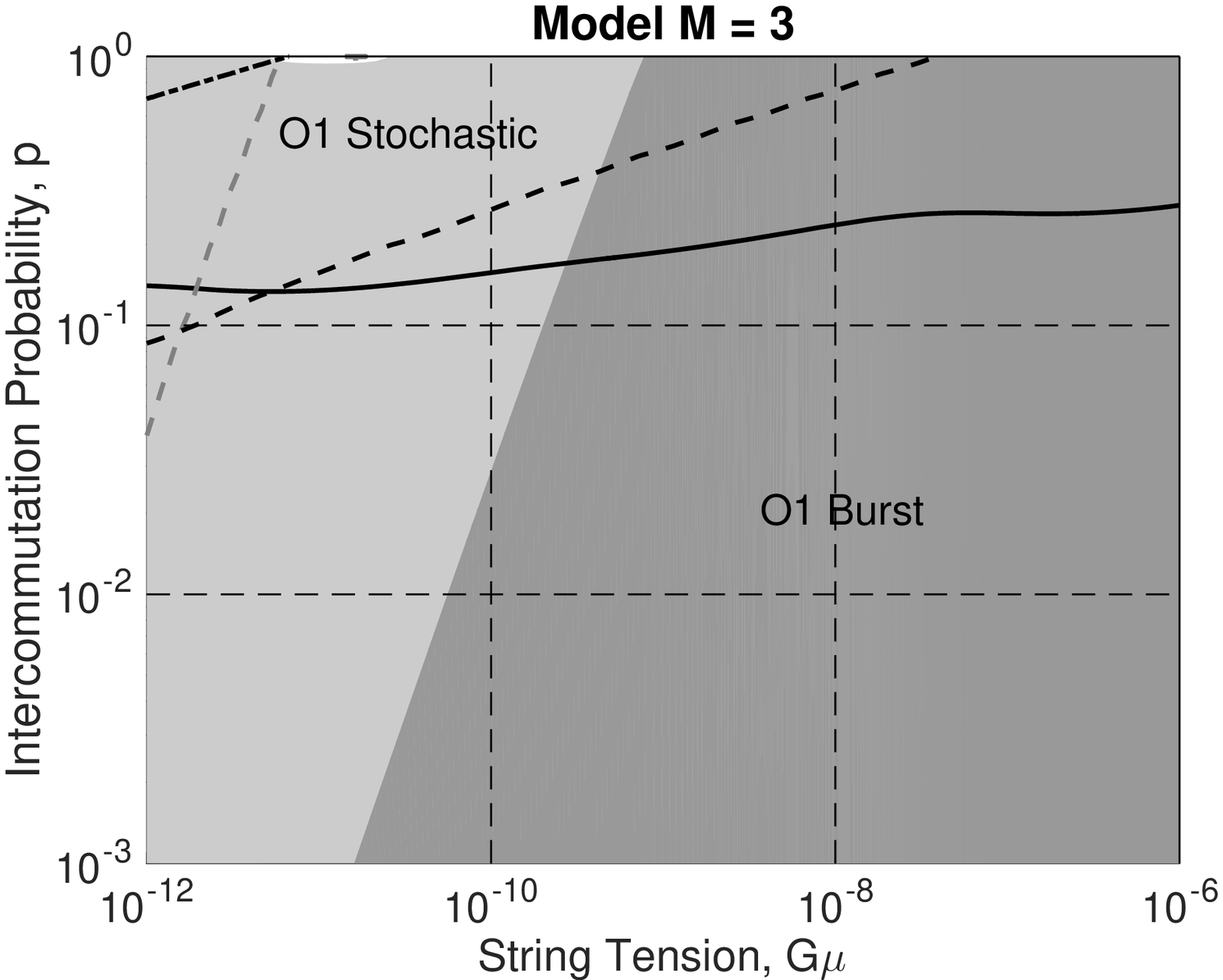}
  \end{subfigure}
  \begin{subfigure}[t]{0.48\textwidth}
    \includegraphics[width=0.6\textwidth,valign=c]{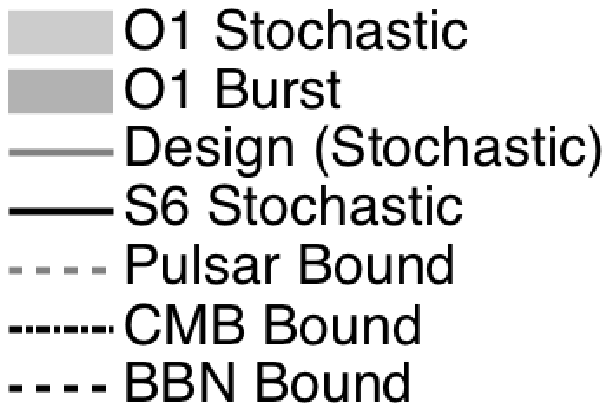}
  \end{subfigure}
  \caption{95\% confidence exclusion regions are shown for three loop 
    distribution models: $M=1$ (top-left), 
    $M=2$ (top-right), and 
    $M=3$ (bottom-left). Shaded regions are excluded by the 
    latest (O1) Advanced LIGO stochastic \cite{TheLIGOScientific:2016dpb} and burst 
    (presented here) measurements. We also show the bounds from the previous 
    LIGO-Virgo stochastic 
    measurement (S6) \cite{Aasi:2014zwg}, from the indirect BBN and CMB bounds 
    \cite{Cyburt:2004yc,Pagano:2015hma}, and from the PTA measurement (Pulsar) \cite{Lasky:2015lej}. 
    Also shown is the projected design sensitivity of 
    the Advanced LIGO and Advanced Virgo experiments (Design, Stochastic) 
    \cite{TheLIGOScientific:2016wyq}. The excluded regions are below the 
    respective curves.}
  \label{fig:contours}
  
\end{figure*}

Fig.~\ref{fig:contours} shows the excluded regions in the parameter spaces of
the three models considered here, based on the stochastic observational constraints
discussed above.

\section{Discussion}\label{sec:conclusion}

The constraints on the cosmic string tension $G\mu$ and intercommutation probability $p$ are shown in Fig.~\ref{fig:contours} for the three loop models under consideration: $M=1$~\cite{Vilenkin:2000jqa, Siemens:2006vk} (top-left), $M=2$~\cite{Blanco-Pillado:2013qja} (top-right) and $M=3$~\cite{Lorenz:2010sm} (bottom-left). We recall that these three models were developed for $p=1$ and, as explained earlier, for smaller intercommutation probability, we used a $1/p$ dependence for the loop distribution.

The bounds resulting from the burst search performed on O1 data are the least constraining. For model 3 and $p=1$, the burst search constraint is $G\mu<8.5 \times 10^{-10}$ at a 95\% confidence level. For models 1 and 2, the burst search can only access superstring models ($p<1$) for which the predicted event rate is larger.

Tighter constraints are obtained when probing the stochastic background of GWs produced by cosmic strings. For model 3, the parameter space studied here is almost entirely excluded by the new constraint derived from the LIGO stochastic O1 analysis. The LIGO stochastic analysis is sensitive to GWs produced in the radiation era. As discussed in Sec.~\ref{sec:models}, in the radiation era, the number of small loops in models 1 and 2 is much smaller than for model 3. When loops are large, the GWs are strongly beamed and the resulting GW detection rate is greatly reduced. As a consequence, experimental bounds using models 1 and 2 are less constraining as can be seen in Fig.~\ref{fig:contours}. For model 1, topological strings ($p=1$) are constrained by $G\mu<5\times 10^{-8}$ with the O1 LIGO stochastic analysis. For model 2, the cosmic string simulation predicts a smaller density of loops and the LIGO constraint is therefore less strict.


In addition to LIGO results, Fig.~\ref{fig:contours} shows limits from pulsar timing experiements, and indirect limits from BBN and CMB data. These experimental results are complementary as they probe different regions of the loop distributions. The CMB and LIGO stochastic bounds apply for the most part to cosmological loops present in the radiation era ($z>3300$). The LIGO burst constraint, although weaker, is sensitive to GWs produced in the matter-era ($z<3300$) from loops which themselves were formed in the radiation era. Constraints from pulsar timing experiments are the most competitive. For topological strings, we get $G\mu<3.8 \times 10^{-12}$, $G\mu<1.5 \times 10^{-11}$ and $G\mu<5.7 \times 10^{-12}$ for models 1, 2 and 3 respectively. However, at nanohertz frequencies, they only probe loops formed in the matter era for very small redshifts corresponding to galactic scales ($z\lesssim 10^{-5}$).


The pulsar bound on string parameters will not improve much in the future as the range of strain amplitudes, $10^{-18}\lesssim h\lesssim 10^{-5}$ (see Fig.~\ref{fig:spectrum} (bottom) and Fig.~\ref{fig:rate2} (right) in Appx.~\ref{appx:tech}), allowed by loop models is already fully explored. The indirect bounds from BBN and CMB data will also be limited by the precision on the $N_{\mathrm{eff}}$ parameter which can be achieved. The sensitivity of Advanced LIGO detectors, however, will further improve in the coming years. In Fig.~\ref{fig:contours} we also report the upper limits the stochastic analysis should achieve with an Advanced LIGO-Virgo detector network working at design sensitivity (see also \cite{Kuroyanagi:2012wm,Kuroyanagi:2012jf}). These will probe most of the parameter space for the three models, and, in particular for models 1 and 3, will surpass all of the current bounds.

\section*{Acknowledgments}
The authors gratefully acknowledge the support of the United States
National Science Foundation (NSF) for the construction and operation of the
LIGO Laboratory and Advanced LIGO as well as the Science and Technology Facilities Council (STFC) of the
United Kingdom, the Max-Planck-Society (MPS), and the State of
Niedersachsen/Germany for support of the construction of Advanced LIGO 
and construction and operation of the GEO600 detector. 
Additional support for Advanced LIGO was provided by the Australian Research Council.
The authors gratefully acknowledge the Italian Istituto Nazionale di Fisica Nucleare (INFN),  
the French Centre National de la Recherche Scientifique (CNRS) and
the Foundation for Fundamental Research on Matter supported by the Netherlands Organisation for Scientific Research, 
for the construction and operation of the Virgo detector
and the creation and support  of the EGO consortium. 
The authors also gratefully acknowledge research support from these agencies as well as by 
the Council of Scientific and Industrial Research of India, 
the Department of Science and Technology, India,
the Science \& Engineering Research Board (SERB), India,
the Ministry of Human Resource Development, India,
the Spanish  Agencia Estatal de Investigaci\'on,
the  Vicepresid\`encia i Conselleria d'Innovaci\'o, Recerca i Turisme and the Conselleria d'Educaci\'o i Universitat del Govern de les Illes Balears,
the Conselleria d'Educaci\'o, Investigaci\'o, Cultura i Esport de la Generalitat Valenciana,
the National Science Centre of Poland,
the Swiss National Science Foundation (SNSF),
the Russian Foundation for Basic Research, 
the Russian Science Foundation,
the European Commission,
the European Regional Development Funds (ERDF),
the Royal Society, 
the Scottish Funding Council, 
the Scottish Universities Physics Alliance, 
the Hungarian Scientific Research Fund (OTKA),
the Lyon Institute of Origins (LIO),
the National Research, Development and Innovation Office Hungary (NKFI), 
the National Research Foundation of Korea,
Industry Canada and the Province of Ontario through the Ministry of Economic Development and Innovation, 
the Natural Science and Engineering Research Council Canada,
the Canadian Institute for Advanced Research,
the Brazilian Ministry of Science, Technology, Innovations, and Communications,
the International Center for Theoretical Physics South American Institute for Fundamental Research (ICTP-SAIFR), 
the Research Grants Council of Hong Kong,
the National Natural Science Foundation of China (NSFC),
the Leverhulme Trust, 
the Research Corporation, 
the Ministry of Science and Technology (MOST), Taiwan
and
the Kavli Foundation.
The authors gratefully acknowledge the support of the NSF, STFC, MPS, INFN, CNRS and the
State of Niedersachsen/Germany for provision of computational resources.

\appendix
\section{$\Lambda$-CDM cosmology}\label{appx:cosmo}

In a $\Lambda$-CDM universe, the Hubble rate at redshift $z$ is given by
\be
H(z) = H_0 {\cal H}(z)~,
\ee
where
\be
{\cal H}(z) = \sqrt{\Omega_\Lambda + \Omega_M(1+z)^3 + \Omega_R {\cal{G}}(z)(1+z)^4 }~.
\ee
We use the latest values of the cosmological parameters \cite{Ade:2015xua},  $H_0=100 h$~km~s$^{-1}$~Mpc$^{-1}$, $h=0.678$, $\Omega_M=0.308$,  $\Omega_R=9.1476\times 10^{-5}$, and $\Omega_\Lambda = 1-\Omega_M-\Omega_R$.
At redshift $z$ in the radiation era, the quantity ${\cal G}(z)$ is directly related to the effective number of degrees of freedom $g_*(z)$ and the effective number of entropic degrees of freedom $g_S(z)$ by \cite{Binetruy:2012ze}
\be
{\cal G}(z) = \frac{g_*(z) g_{S}^{4/3}(0)}{g_*(0) g_{S}^{4/3}(z)}~.
\ee
Following \cite{Binetruy:2012ze} we model it by a piecewise constant function whose value changes at the QCD phase transition ($T=200$MeV), and at electron-positron annihilation ($T=200$keV):
\be
{\cal G}(z) = \left\{\begin{array}{ll} \displaystyle
    1 & {\rm for}\ z<10^9~,\\
    0.83 & {\rm for}\
    10^9<z<2 \times10^{12}~,\\
    0.39 & {\rm for} \ z>2 \times10^{12}~.
\end{array}\right.
\ee
Expressions for cosmic time, proper distance, and proper volume element in term of redshift are given by 
\ba
t(z) = \frac{\varphi_t(z)}{H_0}
&\; \; \; \mathrm{with} \; \; \;  &
\varphi_t(z) = \int_z^\infty \frac{dz'}{{\cal H}(z')(1+z')}~,
\\
r(z) = \frac{\varphi_r(z)} {H_0} &\; \; \mathrm{with} \; \;&
\varphi_r(z) = \int_0^z \frac{dz'}{{\cal H}(z')}~,
\\
dV(z) =\frac{\varphi_V(z)}{H_0^{3}}  dz &\; \; \mathrm{with} \; \;&
\varphi_V(z)= \frac{4\pi \varphi_r^2(z)}{(1+z)^3 {\cal H}(z)}~.
\ea
Asymptotically we have:

\ba
\varphi_t(z\ll1) &\sim& 0.9566  \\
\varphi_t(z\gg1) &\sim& \frac{1}{2\sqrt{\Omega_R{\cal G}(z\gg1)}}z^{-2} \\
\varphi_r(z\ll1) &\sim& z  \\
\varphi_r(z\gg1) &\sim& 3.2086
\ea

\section{Rate of gravitational-wave bursts from cosmic strings}\label{appx:tech}

The detection of GWs from cosmic strings is conditioned by the rate of burst events a cosmic string network generates. In this appendix we outline the rate calculation presented in detail in~\cite{Siemens:2006vk} in a form adapted for the three models under consideration.

The expected rate of GW events, observed at frequency $f$, emitted from a proper volume $dV(z)$ at redshift $z$, in an interval of amplitudes between $A_q$ and $A_q+dA_q$, and for model $M$, is given by
\be
\frac{d^2R^{(M)}_{q}}{dV(z)dA_q}(A_q,z,f) = \frac{1}{1+z} \nu^{(M)}_{q}(A_q,z) \Delta_{q}(A_q,z,f)~,
\label{eq:d2RdVdA}
\ee
where $\Delta_q$ is the fraction of GW events of amplitude $A_q$ that are observable at frequency $f$ and redshift $z$. Since cusps emit GW bursts in a cone of solid angle $d\Omega \sim \pi \theta_m^2$ (where $\theta_m$ is given in Eq.~\ref{thetam}) and kinks into a fan-shaped set of directions in a solid angle $ d\Omega \sim 2\pi \theta_m$, one finds
\ba
\Delta_{q}(A_q,z,f)&\sim & \left(\frac{\theta_m(\ell,z,f)}{2}\right)^{3(2-q)}
\nonumber
\\
&& \qquad \times \Theta\big(1-\theta_m(z,f,\ell) \big)
\label{eq:delta}
\ea
where $\ell=\ell(A_q,z)$ is obtained by inverting Eq.~\ref{eq:A}.

The number of cusp/kink features per unit space-time volume on loops with sizes between $\ell$ and $\ell+d\ell$ is given by
\be
\nu_q^{(M)}(\ell,z)d\ell = \frac{2}{\ell} N_{q} n^{(M)}(\ell,t(z))d\ell~,
\label{eq:nuoflz}
\ee
where $N_{q}$ is the number of cusps/kinks per oscillation period. Using Eq.~\ref{eq:A} to change variables from $\ell$ to $A_q$ gives
\ba
\nu_q^{(M)}(A_q,z)dA_q &=& \nu_q^{(M)}(\ell(A_q,z),z)\frac{d\ell}{dA_q}dA_q \nonumber \\
&=& \nu_q^{(M)}(\ell(A_q,z),z)\frac{\ell(A_q,z)}{(2-q)A_q}dA_q.
\label{eq:nuofAz}
\ea

Injecting the loop distribution of model $M$, ${\cal F}^{(M)}$, Eq.~\ref{eq:d2RdVdA} becomes
\ba
\frac{d^2R^{(M)}_{q}}{dzdA_q}(A_q,z,f) &=& \frac{2N_qH_0^{-3}\varphi_V(z)}{(2-q)(1+z)A_qt^4(z)} \nonumber \\
&& \times {\cal F}^{(M)}\left(\frac{\ell(A_q,z)}{t(z)},t(z)\right)  \nonumber \\
&& \times \Delta_{q}(A_q,z,f).
\label{eq:d2RdzdA}
\ea
Alternatively, the rate can also be parameterized by the strain amplitude using Eq.~\ref{eq:wf}:
\ba
\frac{d^2R^{(M)}_{q}}{dzdh}(h,z,f) &=& \frac{2N_qH_0^{-3}\varphi_V(z)}{(2-q)(1+z)ht^4(z)} \nonumber \\
&& \times {\cal F}^{(M)}\left(\frac{\ell(hf^q,z)}{t(z)},t(z)\right)  \nonumber \\
&& \times \Delta_{q}(hf^q,z,f).
\label{eq:d2Rdzdh2}
\ea

The rate of GWs given in Eq.~\ref{eq:d2Rdzdh2} is marginalized over the strain amplitude and the redshift to compute the GW stochastic background (see Eq.~\ref{eq:spec}).
The strain amplitude range is limited by two physical conditions: firstly the beaming angle must satisfy $\theta_m<1$, secondly, in all three models, there is an upper bound for the loop size, $\gamma_\mathrm{max}$. These conditions straightforwardly impose (see Eq.~\ref{eq:wf} and Eq.~\ref{eq:A}) that $h_\mathrm{min}(z) < h < h_\mathrm{max}(z)$, where
\ba
h_\mathrm{min}(z)&=&\frac{G\mu H_0}{f^2(1+z)\varphi_r(z)}\label{eq:numlim1}\\
h_\mathrm{max}(z) &=& \frac{(\gamma_\mathrm{max}\varphi_t(z))^{2-q}G\mu}{H_0^{q+1}(1+z)^{q-1}f^q\varphi_r(z)} \label{eq:numlim2}.
\ea
In turn, the condition $h_\mathrm{min}(z) \leq h_\mathrm{max}(z)$ fixes the upper limit on the redshift, $z_\mathrm{max}$. Finally, the overall GW rate is obtained by calculating the double integral:
\be
R^{(M)}_{q}=\int_{0}^{z_\mathrm{max}}dz\int_{h_\mathrm{min}(z)}^{h_\mathrm{max}(z)}dh\frac{d^2R^{(M)}_{q}}{dzdh}(h,z,f) \label{eq:avgrate}.
\ee

\begin{figure*}
  \center
  \epsfig{width=7.7cm, file=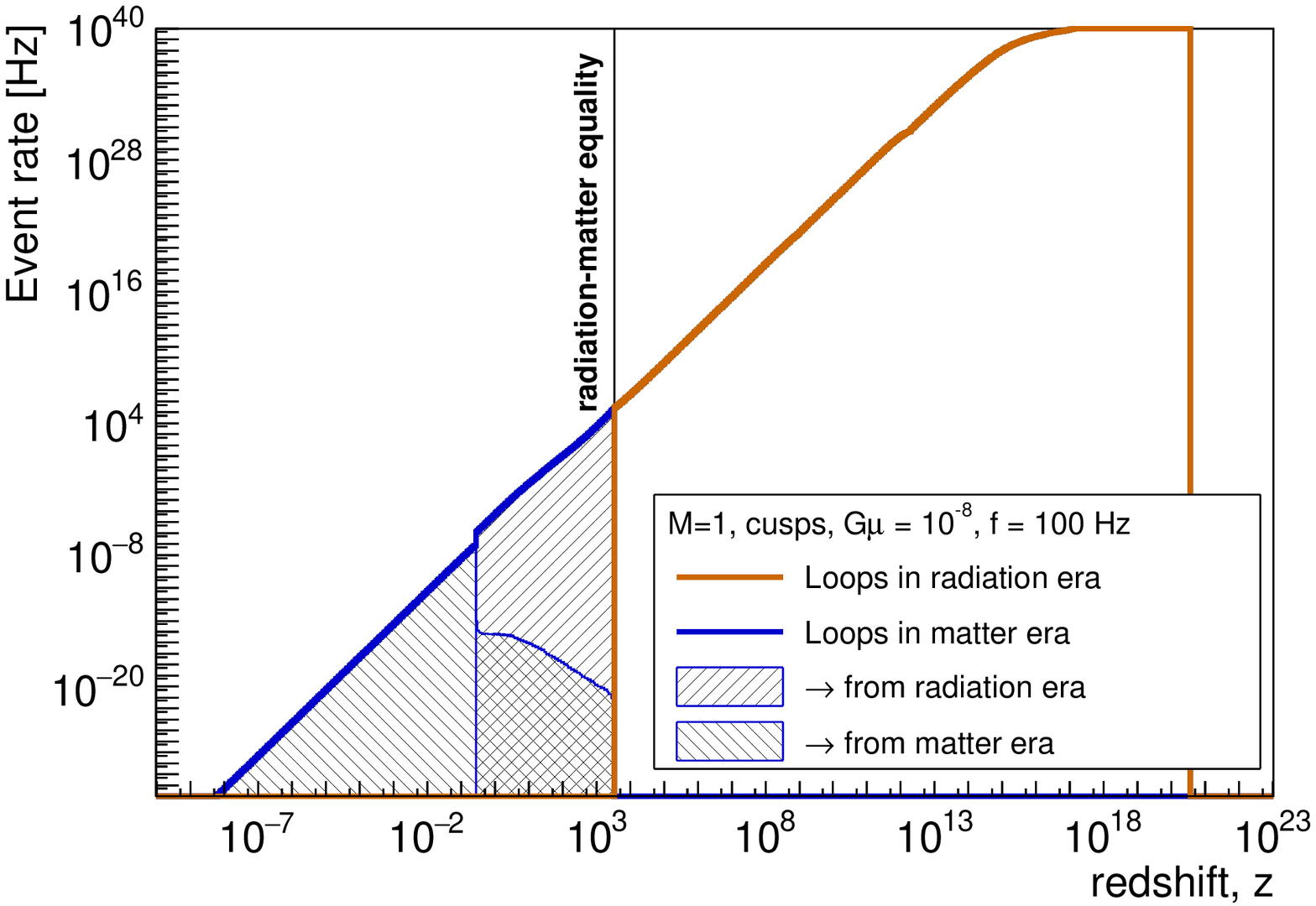} \epsfig{width=7.7cm, file=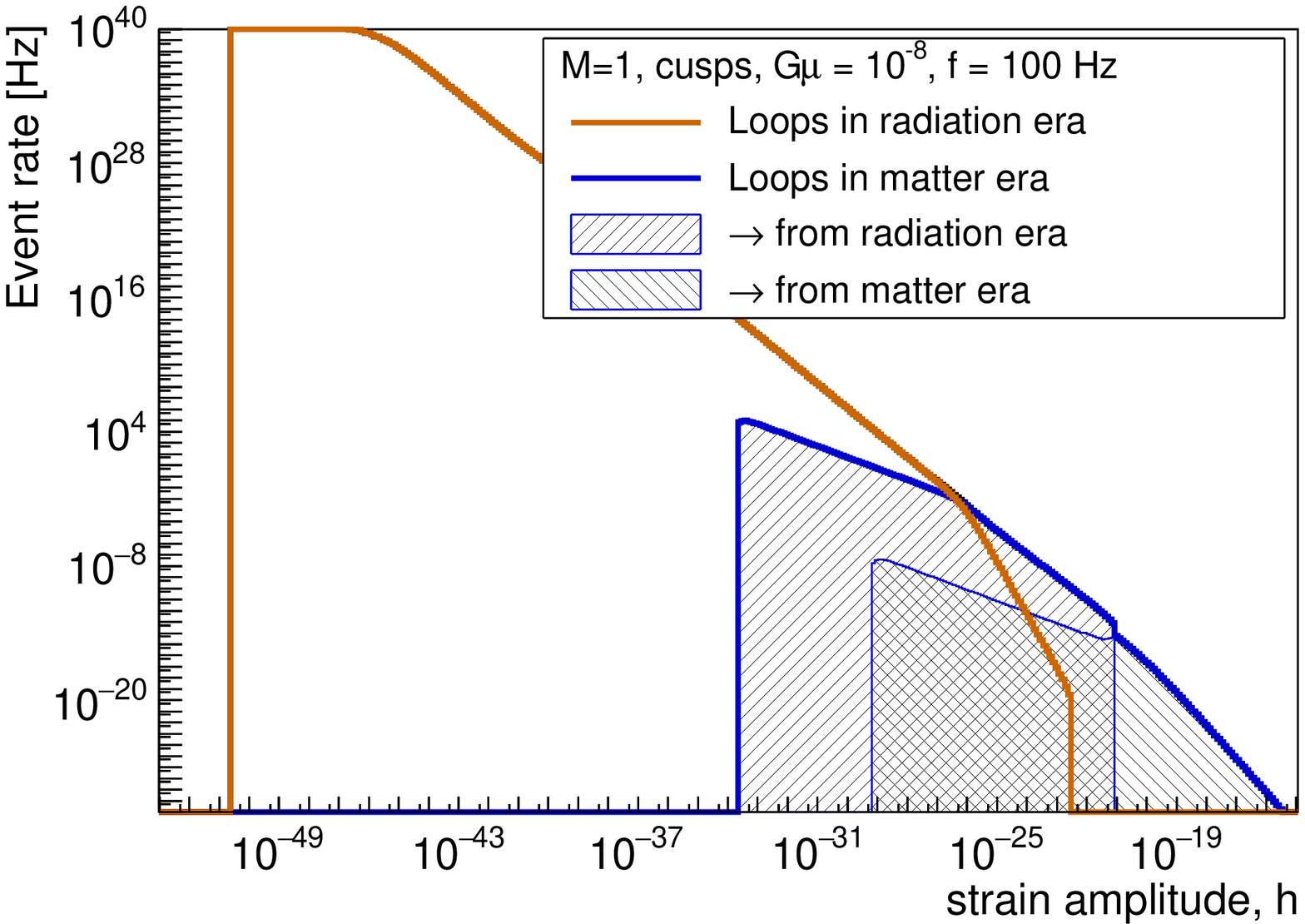} \\
  \epsfig{width=7.7cm, file=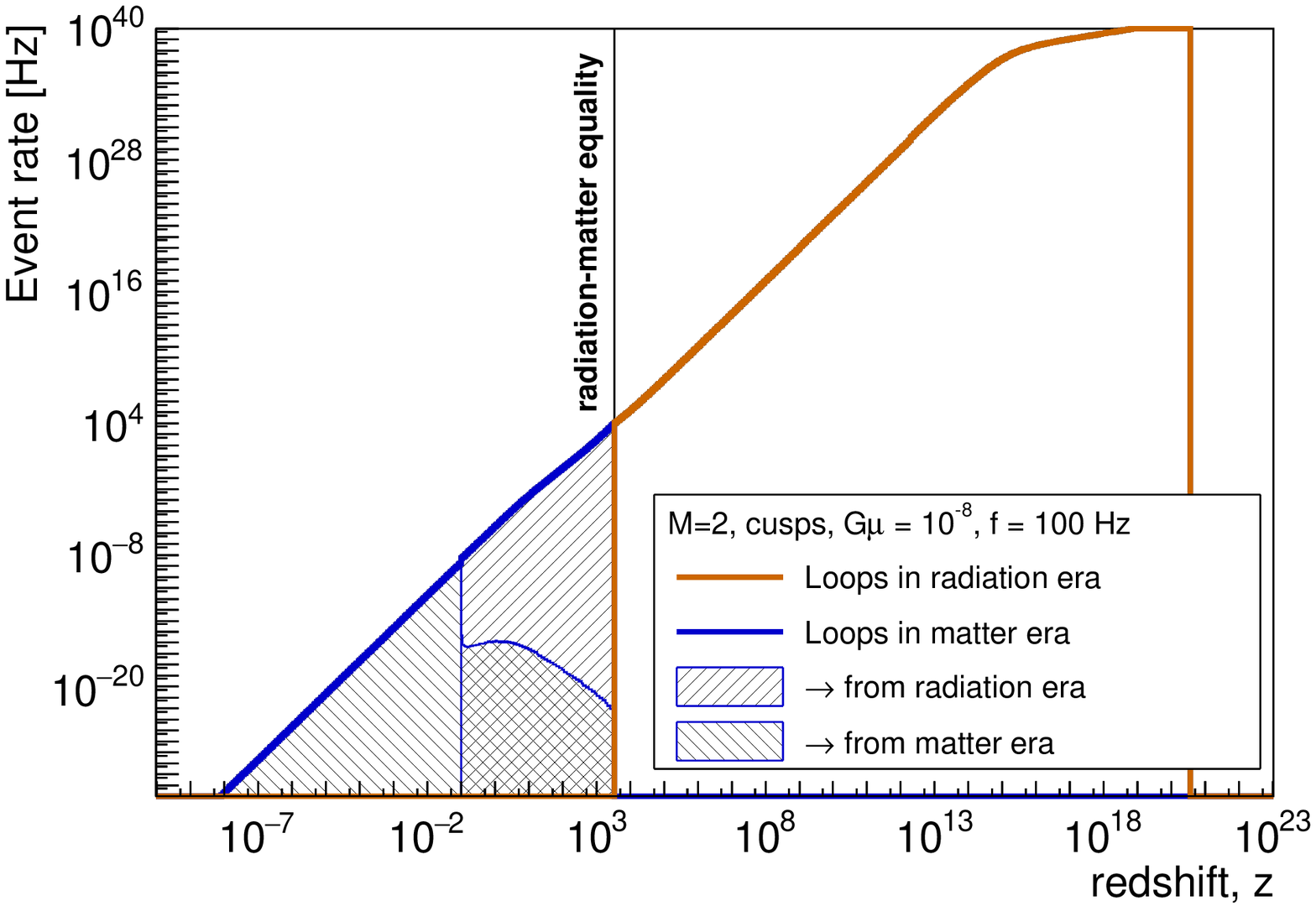} \epsfig{width=7.7cm, file=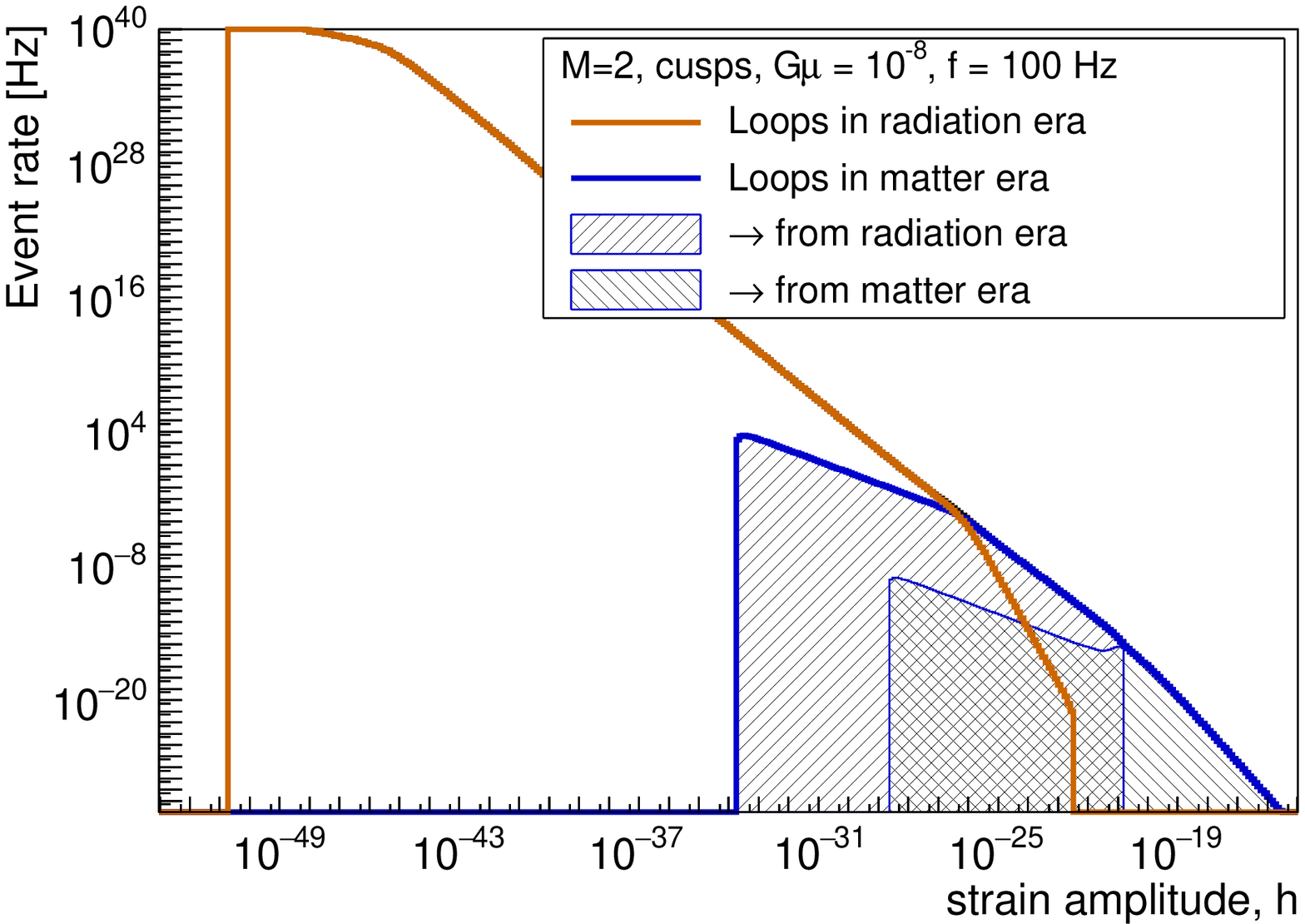} \\
  \epsfig{width=7.7cm, file=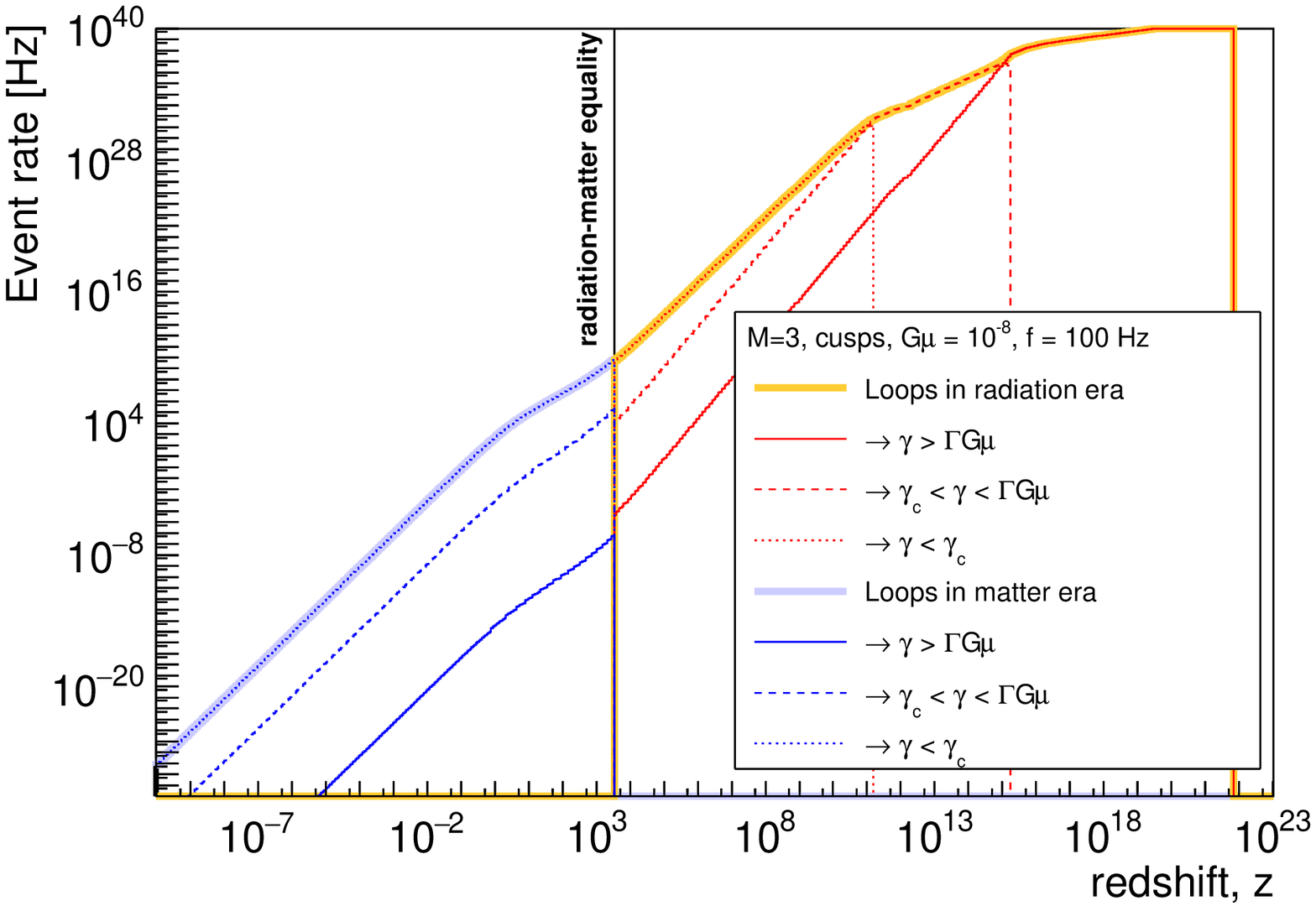} \epsfig{width=7.7cm, file=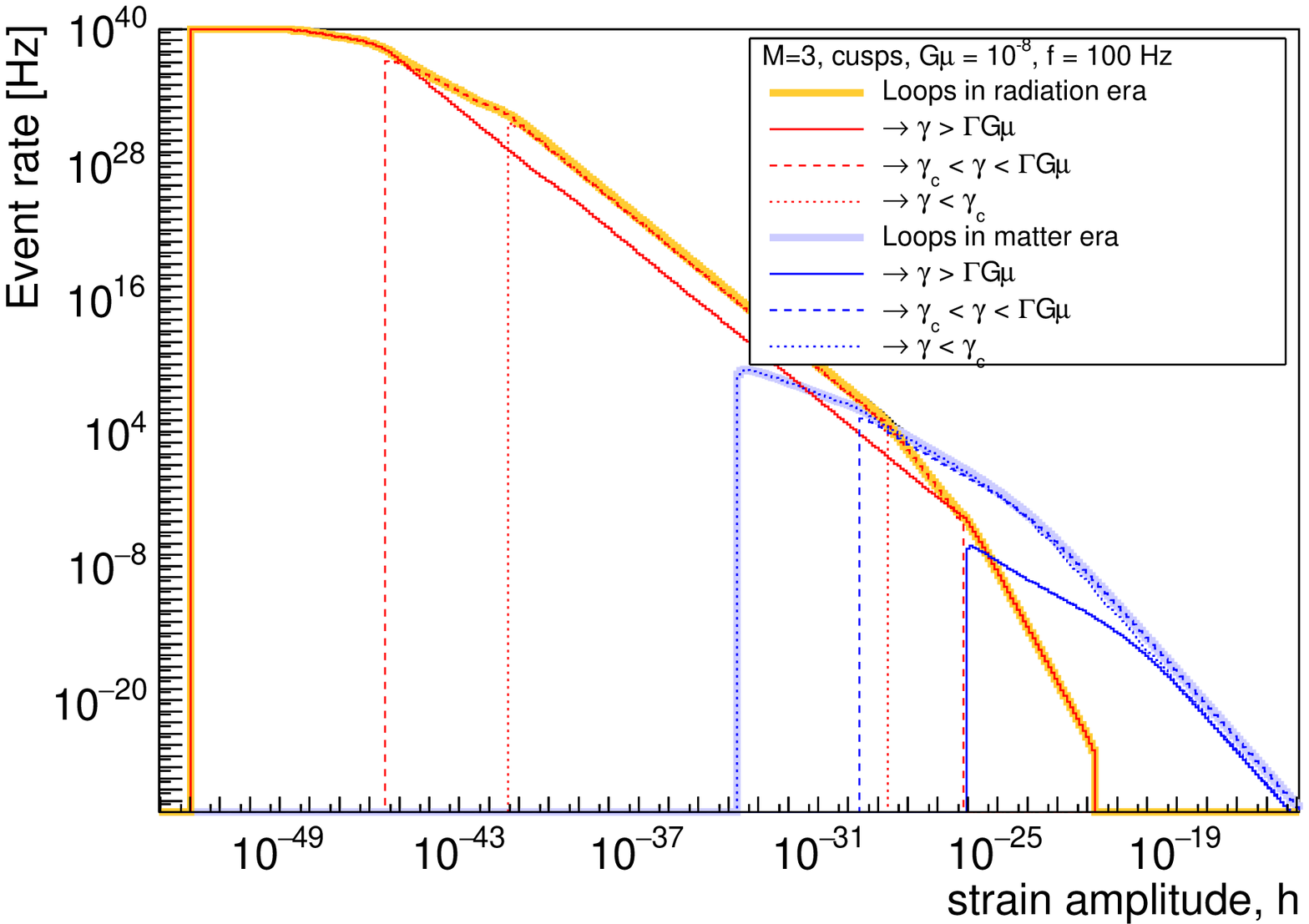} \\
  \caption{GW event rate predicted by models $M=1$ (top row), $M=2$ (middle row) and $M=3$ (bottom row) and averaged over either $h$ (left column) or $z$ (right column). The string tension and the wave frequency are fixed to $10^{-8}$  and to $100$~Hz respectively. For models 1 and 2, we separated the contributions from loops in the radiation ($z>3366$) and matter ($z<3366$) eras. Additionally, for loops in the matter era, we separated the effect of loops produced in the matter era from the ones produced in the radiation era (Eq.~\ref{dist-rad}, Eq.~\ref{dist-1} and Eq.~\ref{dist-2}). For model 3, the effect of the three loop size regimes is shown (Eq.~\ref{eq:mod3_1}, Eq.~\ref{eq:mod3_2} and Eq.~\ref{eq:mod3_3}).}
  \label{fig:rate1}
\end{figure*}

For illustration, we fix $f=100$~Hz and $G\mu=10^{-8}$ and we average the GW rate for cusps over either $h$ (left-hand column of Fig.~\ref{fig:rate1}) or $z$ (right-hand column of  Fig.~\ref{fig:rate1}) for models $M=\{1,2,3\}$. The contributions from loops in the matter (blue curve) and radiation (red curve) eras are also presented. For loops in the matter era, we separated the effect of loops produced in the radiation era from loops produced in the matter era.

We first observe that all models have the same general dependence on redshift and strain amplitude: high-amplitude GWs are produced in the matter era with a low rate while weak GWs are produced in the radiation era with a high rate. Models differ in the absolute rate of GWs they predict: $R^{(M)}_\mathrm{cusps}(h=10^{-23})=1.1\times 10^{-9}$~Hz, $1.2\times 10^{-10}$~Hz and $1.0\times 10^{-6}$~Hz for models 1, 2 and 3 respectively. As noted in Sec.~\ref{sec:models}, in model 3, small loops are copiously present at all times. Indeed, for model 3, the small loop contribution to the GW rate dominates for $h\apprge 10^{-45}$ and $z\apprle 10^{15}$, while for models 1 and 2 it is negligible.

\begin{figure*}
  \center
  \epsfig{width=7.7cm, file=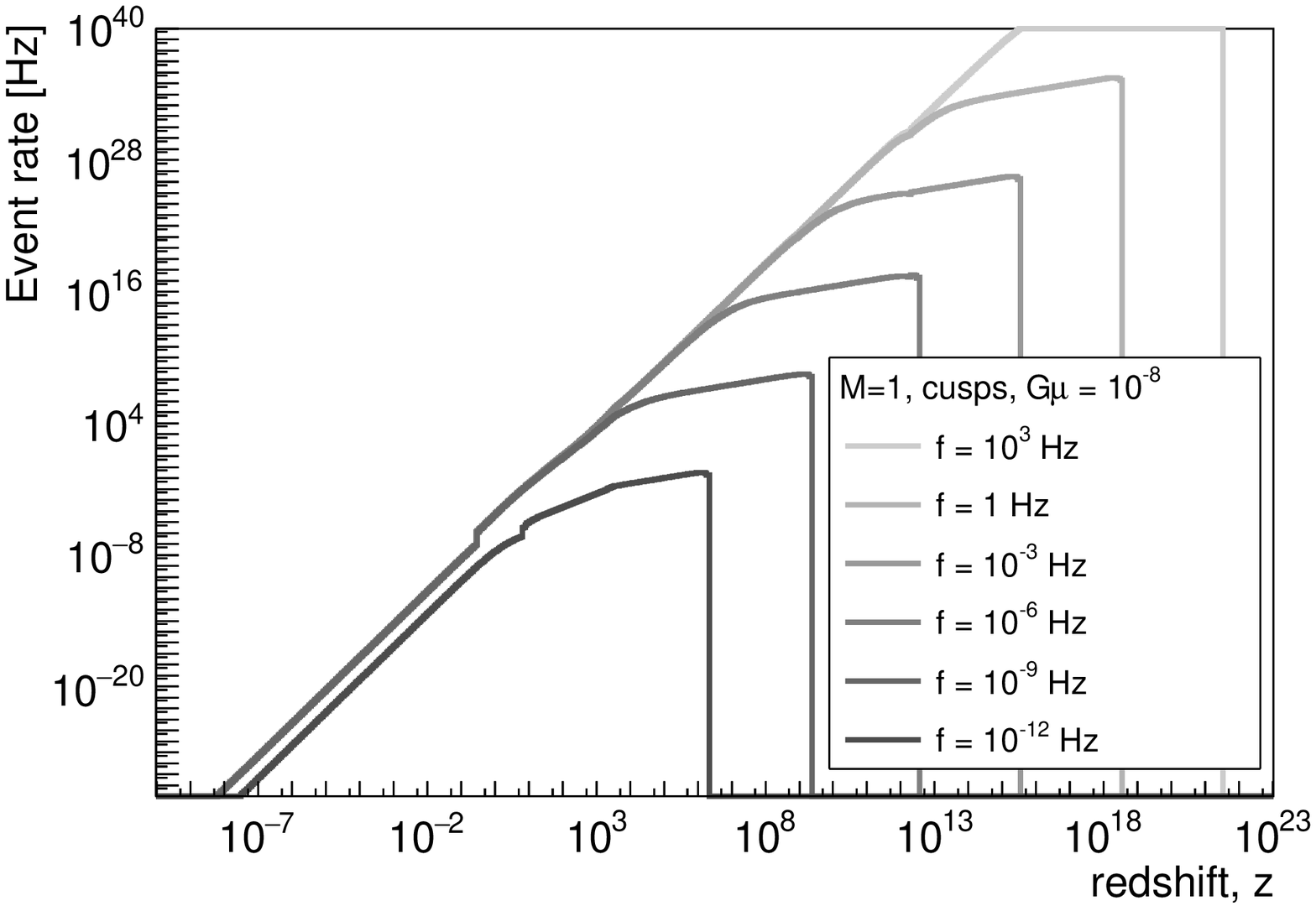} \epsfig{width=7.7cm, file=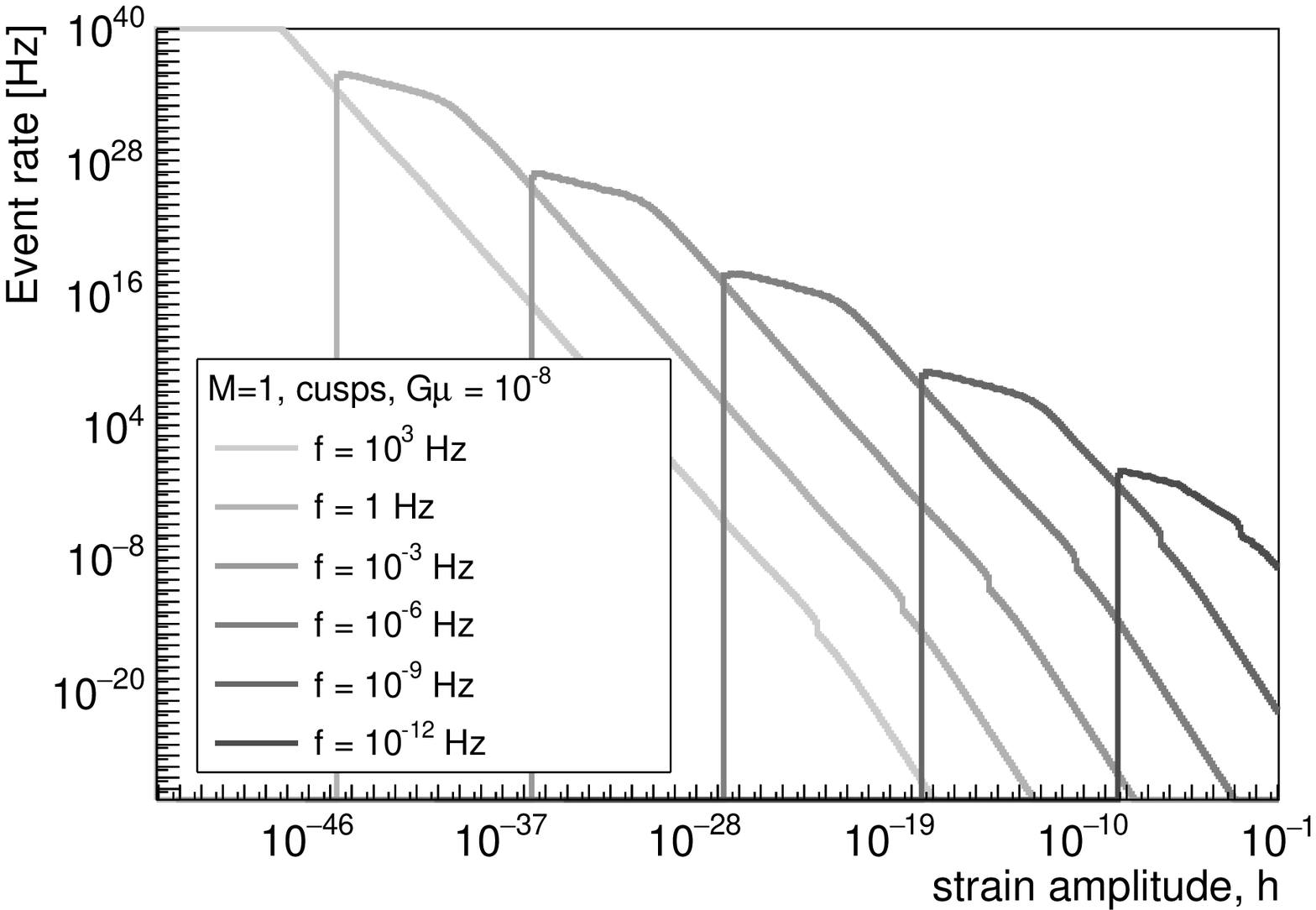}
  \caption{GW event rate predicted by model $M=1$ for different frequencies and averaged over either $h$ (left column) or $z$ (right column).}
  \label{fig:rate2}
\end{figure*}

In Fig.~\ref{fig:rate2}, the effect of the wave frequency $f$ is studied ($M=1$ only). Loops in the radiation era tend to produce high-frequency GWs while low-frequency waves are emitted in the matter era. The event rates presented in Fig.~\ref{fig:rate2} condition the detectability of GWs from cosmic strings using experimental data sets.

\bibliographystyle{apsrev}
\bibliography{references}

\end{document}